\documentclass[iop,twocolumn]{emulateapj-rtx4}

\usepackage{graphicx}  
\usepackage{dcolumn}   
\usepackage{bm}        
\usepackage{amsfonts,amsmath,amssymb,mathrsfs}
\usepackage{color}
\usepackage{hyperref}

\usepackage{natbib,times}
\definecolor{orange}{rgb}{1,0.5,0}

\newcommand{\cf}{\textit{cf.}~}
\newcommand{\ie}{\textit{i.e.}~}
\newcommand{\eg}{\textit{e.g.}~}

\shorttitle{Poloidal-Field Instability in Magnetized Relativistic Stars} 
\shortauthors{R. Ciolfi and L. Rezzolla}

\begin{document}

\title{Poloidal-Field Instability in Magnetized Relativistic Stars}

\author{Riccardo Ciolfi\altaffilmark{1} and Luciano Rezzolla\altaffilmark{1,2}}

\altaffiltext{1}{Max-Planck-Institut f\"ur Gravitationsphysik,
  Albert-Einstein-Institut, Potsdam, Germany}

\altaffiltext{2}{Department of Physics and Astronomy, Louisiana State
  University, Baton Rouge, Louisiana, USA}

\begin{abstract}
We investigate the instability of purely poloidal magnetic fields in
nonrotating neutron stars by means of three-dimensional
general-relativistic magnetohydrodynamics simulations, extending the
work presented in \citet{Ciolfi2011}. Our aim is to draw a clear
picture of the dynamics associated with the instability and to study the
final configuration reached by the system, thus obtaining indications
on possible equilibria in a magnetized neutron star. Furthermore,
since the internal rearrangement of magnetic fields is a highly
dynamical process, which has been suggested to be behind magnetar
giant flares, our simulations can provide a realistic estimate of the
electromagnetic and gravitational-wave emission which should accompany
the flare event. Our main findings are the following: (i) the initial
development of the instability meets all the expectations of
perturbative studies in terms of the location of the seed of the
instability, the timescale for its growth and the generation of a
toroidal component; (ii) in the subsequent nonlinear reorganization of
the system, $\sim 90$\% of magnetic energy is lost in few Alfv\'en
timescales mainly through electromagnetic emission, and further
decreases on a much longer timescale; (iii) all stellar models tend to 
achieve a significant amount of magnetic helicity and the 
equipartition of energy between poloidal and toroidal magnetic fields, and 
evolve to a new configuration which does not show a subsequent instability on
dynamical or Alfv\'en timescales; (iv) the electromagnetic emission
matches the duration of the initial burst in luminosity observed in
giant flares, giving support to the internal rearrangement scenario;
(v) only a small fraction of the energy released during the process is
converted into $f$-mode oscillations and in the consequent
gravitational-wave emission, thus resulting in very low chances of
detecting this signal with present and near future ground based
detectors.
\end{abstract}
		
	

\section{Introduction}

Neutron stars (NSs) are endowed with very intense, long-lived,
large-scale magnetic fields, reaching strengths which are estimated to
be of the order of $10^{13}$ G at the magnetic pole for ordinary NSs,
and around $10^{15}$ G in the case of magnetars. Such extreme magnetic
fields play a crucial role in the physics of NSs, affecting their
structure and evolution. They are involved in the processes through
which NSs are observed, like the pulsar magnetic dipole radiation and
the magnetically-powered burst activity of magnetars, and they have
been recently recognised as essential to explain the quasi-periodic
oscillations detected in the aftermath of magnetar giant flares [see,
  \eg, \citet{Gabler2012} and references therein]. Moreover, they are
responsible for deformations which may cause a significant emission of
gravitational waves \citep{Bonazzola1996,Cutler2002} and precession
\citep{Wasserman2003}, they influence the thermal evolution of the
star \citep{Pons2009}, to list a few.

All these processes depend on the magnetic field configuration inside
the NSs, whose geometry is basically unknown. From observations of the
spindown in pulsars the exterior magnetic field appears
to be purely poloidal and mainly dipolar, but substantially different
internal geometries can reproduce this external appearance. The
importance of obtaining such information has motivated a significant
effort in studying possible equilibrium models of magnetized NSs, at
first with simple geometries, \eg purely poloidal or purely toroidal
fields, and recently with mixed poloidal-toroidal fields. The latest
models, built in Newtonian and general-relativistic framework, include
\citet{Tomimura2005, Lander:2009, Lander2012, Ciolfi2009, Ciolfi2010, 
Fuji2012}, where the so-called `twisted-torus' configuration is considered. 
This particular geometry has been found as a result of the evolution of 
initial random fields in Newtonian magnetohydrodynamic (MHD) 
simulations by \citet{BraithNord2006}.

Once a magnetic-field geometry is chosen, building a corresponding
equilibrium configuration is not sufficient to assess whether this
represents a good description of the NS interior. The magnetic field,
in fact, should also be long-lived and thus stable on timescales which
are much longer than the dynamical timescale. Assessing the stability
of a given magnetic field configuration is not trivial and most of the
work done on the subject concerns simple field geometries and
nonrotating stars. Until very recently, the problem has been only
addressed with a perturbative analytic approach. These calculations
established that both a purely poloidal field and a purely toroidal
field are unstable in nonrotating stars, giving important predictions
about the onset of the instability, but they could not predict the
following evolution of the system. Only recently, thanks to the
progress in numerical simulations, it has become possible to study
these hydromagnetic instabilities by performing the fully
three-dimensional (3D) MHD evolutions of magnetized relativistic
stars. These simulations represent a very powerful tool, allowing to
confirm the predicted features of the instability and to obtain
information about the nonlinear dynamics of the process. In addition,
the end-state of simulations can provide important hints about the
preferred magnetic field configuration in magnetized stars.

There is an additional and important motivation for studying hydromagnetic 
instabilities in NSs. The induced global rearrangement of
magnetic fields is a violent, strongly dynamical process, and soon after the 
magnetar model was proposed \citep{DT92}, this kind of process was 
suggested as a trigger mechanism of giant flares \citep{TD95,TD2001}. 
Nowadays, this internal rearrangement scenario represents
one of the leading models to explain the phenomenology observed in
magnetars, the other one involving a large-scale rearrangement of
magnetic fields in the magnetosphere surrounding the 
star~\citep{Lyutikov2003,Lyutikov2006,Gill2010}. 
Since both mechanisms can be present in a giant flare, the 
main question becomes whether most of the magnetic energy 
powering the flare is stored inside the star or in its exterior 
magnetosphere. The basic tests on these models rely on
the comparison of the predicted timescales and energies involved with
the giant flare observations. Hence, determining self-consistently the
dynamics associated to this kind of instability can provide
important hints on the underlying mechanism.

Moreover, magnetar flares (and in particular giant flares) are likely
to be accompanied by a significant excitation of NS oscillations, in
particular in the $f$-mode, which can then lead to a strong emission
of gravitational waves (GWs). This possibility has motivated recent
searches for GWs in connection to magnetar flares, published by the
LIGO and Virgo collaboration [see, \eg
  \citet{ligo-virgo2011}]. Semi-analytic efforts have been devoted to
establishing the maximum amount of magnetic energy released in a
magnetar flare, which, in turn, provides an upper limit on the energy
emitted in GWs \citep{Ioka01,Corsi2011}. These upper limits are based
on analytical calculations and simplified models, and can only provide
rough, order-of-magnitude estimates. The assumption that all the
available energy (which is at most of the order of the total magnetic
energy) is converted into GWs, leads to the optimistic conclusion that
the signal would be detectable with the next-generation ground-based
detectors \citep{Corsi2011}. This result has been questioned in
\citet{Levin:2011}, where a simple perturbative analysis is employed
to show that only a small fraction of the magnetic energy involved in
a giant flare event can be actually be converted into $f$-mode
oscillations and that the consequent GW emission is expected to be
very weak. Again, referring to the internal magnetic field
rearrangement scenario of giant flares, MHD simulations of
hydromagnetic instabilities can provide a realistic picture of the GW
signal produced, together with estimates of the fraction of energy
which can be pumped into the $f$-mode and of the signal detectability.

In this work we focus on the instability of purely poloidal fields,
with the goal of shedding some light on all of the points made
above\footnote{The instability of purely-toroidal fields has several
  analogies with the one considered here for purely-poloidal fields
  and has been investigated by \citet{Kiuchi:2008,Kiuchi:2011}.}. Two
parallel works back in the '70s~\citep{Markey1973,Wright1973} found
that poloidal fields suffer from the so-called ``Tayler'' or ``kink''
instability, which manifests itself firstly in the neighbourhood of
the neutral line. This instability was recently studied with Newtonian
numerical simulations in the linear regime \citep{Lander:2011}, or
with nonlinear evolutions for a simplified model of newly born NS
\citep{Geppert2006}, and in the case of main-sequence stars
\citep{Braithwaite2007}. The first 3D general-relativistic MHD
simulations of the poloidal field instability in NSs were presented
only last year, in two parallel works
\citep{Lasky2011,Ciolfi2011}. These studies reported similar results,
despite some substantial difference in the approach (in particular in
the evolution of magnetic fields outside the star), essentially
confirming all of the analytic predictions on the instability, and
providing some first hints about the nonlinear rearrangement of the
system. In addition, \citet{Ciolfi2011} presented the first examples
of gravitational waveforms triggered by the instability, which were
subsequently considered more systematically by \citet{Zink:2011} for
nonrotating stars and by \citet{Lasky2012} for rotating NSs. In this
paper we extend the work presented in \citet{Ciolfi2011}, presenting
additional information on the numerical infrastructure used and
considering the instability-driven evolution of a series of
nonrotating NSs endowed with purely poloidal magnetic fields of
different strength.

The organization of the paper is as follows. In Sect.~\ref{setup} we
reconsider the setup of the system, improving in particular our
treatment of the atmosphere. Within the new setup, we confirm that our
evolutions meet all the expectations on the onset of the instability,
in agreement with the previous perturbative studies. In
Sect.~\ref{evolution} we examine in more detail the nonlinear
rearrangement of the system, performing much longer simulations and
gaining new substantial insight on the final state reached by the
system. Section~\ref{endstate} is dedicated to a discussion of the
implications for the most-likely magnetic-field configurations in
magnetized NSs and of the role played by magnetic helicity. We also
study the emission properties of the system, relevant for the internal
field rearrangement scenario of giant flares, estimating in
Sect.~\ref{ememission} the timescale of the process and its
electromagnetic luminosity, and discussing the detectability of the
GW signal in Sect.~\ref{gravitationalwaves}. Both for electromagnetic
and GW emissions, our estimates rely on a good agreement with the
dependence on the magnetic field strength expected in the perturbative
limit of weak magnetic fields, which allows us to extrapolate our
results to lower and more realistic values than those actually
considered in the simulations. Our conclusions are finally presented
in Sect.~\ref{sec:conclusions}. Unless specified differently, we adopt
units in which $c=1$, $G=1$.


\section{Physical system and numerical setup}\label{setup}

Our physical system of interest is represented by a nonrotating
isolated neutron star surrounded by vacuum, initially endowed with a
purely poloidal magnetic field permeating the star and extending to
the exterior. The initial configurations are fully-relativistic
self-consistent solutions generated with the multi-domain
spectral-method code \texttt{LORENE}, developed at the Observatoire de
Paris-Meudon \citep{Bocquet1995}. The stars are modeled as composed of
a barotropic fluid obeying a polytropic equation of state $p\equiv
K\rho^{\Gamma}$, with $\Gamma=2$ and $K=100$. The reference
unmagnetized star has a gravitational mass of $1.4\,M_\odot$ 
and a radius of 12.2 km. Magnetic field
strengths at the pole vary in the range\footnote{For simplicity,
  hereafter we will use a more compact notation for magnetic-field
  strengths in scales of $10^{16}\,{\rm G}$, \ie \hbox{$B_N = N \times
    10^{16}\,{\rm G}$}.} $B_{\rm p}=B_{1.0}-B_{9.5} \equiv (1-9.5) \times
10^{16}\,{\rm G}$. A stronger magnetic field shortens the evolution
timescale of the system, and the above choice makes our simulations
computationally feasible. On the other hand, as we discuss in
Sect.~\ref{results}, most of our results can be extrapolated back to
smaller (and more realistic) magnetic field strengths.

We perform fully 3D general-relativistic MHD simulations of the system
adopting the Cowling approximation, \ie~neglecting changes in the
spacetime metric. Evolutions are obtained with the \texttt{WhiskyMHD}
code. Most of the details on our mathematical and numerical setup are
discussed in depth in~\citet{Pollney:2007ss, Giacomazzo:2007ti,
  Giacomazzo:2009mp, Giacomazzo:2011}. One aspect worth stressing is
that, to guarantee the divergence-free character of the MHD equations,
we use as an evolution variable the vector potential instead of the
magnetic field, as described in \citet{Giacomazzo:2011}. Our standard
grid setup consists of three refinement levels using the
\texttt{Carpet} driver \citep{Schnetter-etal-03b}, with the finest one
covering the entire star and having resolution $h/M_\odot=0.17$ ($\sim 
250$ m). Our computational box extends to $\pm \,54\,M_\odot \sim 80$
km and we impose no symmetries.

To shorten the time for the development of the instability and thus
reduce computational costs, a small perturbation is added to the
initial velocity of the fluid. In particular, we add a
$\theta$-component of fluid velocity in the region surrounding the
neutral line, where the instability is expected to be triggered, and
with $m=2$ azimuthal distribution. The strength of the perturbation
corresponds to relative changes in the magnetic field of the order of
$10^{-3}$. We have verified that the instability occurs even in
absence of initial perturbations and that no appreciable differences
(besides that of reducing computational costs) are introduced in the
dynamics by the perturbation.

In our numerical setup, special care is paid to the treatment of the
atmosphere. We recall that, as customary in finite-volume
relativistic hydrodynamics simulations, our star is surrounded by a
fluid at much lower densities (\ie the ``atmosphere''), obtained by
imposing a minimum rest-mass density [see, \eg \citet{Baiotti04} and
  \cite{Baiotti08}]. In the atmospheric region, where the density is
equal to the imposed floor value, we set the fluid velocity to zero to
avoid the spurious accretion of the atmosphere onto the star. Within
the common assumption of ideal MHD, also adopted in the
\texttt{WhiskyMHD} code, this prescription would imply that magnetic
fields do not evolve in the atmosphere, which represents a serious
limitation from both the physical and numerical point of view. In
particular, this rapidly leads to errors at the stellar surface and,
for the magnetic field strengths we consider, ends the simulations
prematurely.

\begin{figure}
  \begin{center}
     \includegraphics[angle=0,width=8.0cm]{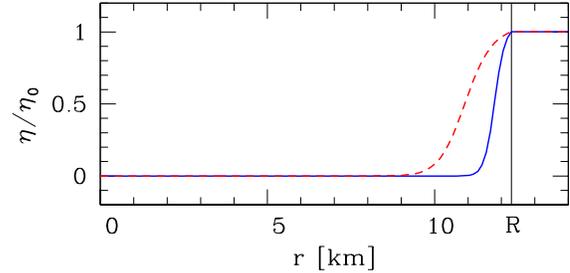}
  \end{center} 
  \caption{Resistivity profiles along the radial direction for our
    fiducial simulation with $B_{\rm p}=B_{6.5}$. The two lines
    refer to the choice of a wide resistive transition layer (red
    dashed line) and corresponding to the one adopted in
    \citet{Ciolfi2011}, or of a thin layer (blue solid line). The
    vertical line marks the radius of the star $R$.}
\label{fig:diffusion}
\end{figure}

A self-consistent solution to this problem would require the
implementation in general-relativity of the equations of resistive
MHD, along the lines of the work reported by~\citet{Palenzuela:2008sf}
and which has seen recent progress in the work
of~\citet{Kiki2012}. Lacking for the time being this more systematic
solution, a reasonable first approximation can be obtained by adding a
resistive term to the evolution equation for the vector potential, \ie
\begin{equation}
\label{a}
\partial_t \vec{A} = 
\vec{\tilde{v}} \times \vec{B} + \eta \Delta \vec{A} \,,
\end{equation}
where $\vec{\tilde{v}}=\alpha \vec{v}-\vec{\beta}$ [see 
\citet{Anton06} for the notation], $\eta$ is the scalar resistivity 
and, for simplicity, we take $\Delta$ to represent the Laplacian 
in a flat spacetime. 
We impose that resistivity is zero inside the star (thus retaining an ideal-MHD
behaviour), up to a transition layer, where it continuously increases
towards the atmospheric value $\eta_0$. This is shown in
Fig.~\ref{fig:diffusion}, where we plot the initial resistivity
profile along the radial direction for our fiducial simulation with
$B_{\rm p}= B_{6.5}$. The plot shows two different radial profiles,
where we change the width of the transition region inside the star,
whose surface $r=R$ is indicated with a vertical solid line. As we will
discuss below, the two choices have a different impact on the
overall results. 
The resistivity profile is set to be an explicit
function of the rest-mass density, namely $\eta(\rho)/\eta_0=g(\rho)$, 
where $g$ is a Fermi-like function, with $g=1$ in the atmosphere and $g=0$ 
inside the stellar core. In this way, the
resistivity will mimic the evolution of the rest-mass density,
following the stellar surface in its evolution.

Since the velocity is set to zero outside the star, eq.~(\ref{a})
reduces there to $\partial_t \vec{A} = \eta_0 \Delta \vec{A}$. A few
remarks are worth doing about this limit and we start by considering
the special relativistic case for simplicity. Maxwell's equations in
vacuum are simple wave equations, \eg $\partial_t^{2} \vec{B} = c^2
\Delta \vec{B}$, so that a non-zero Laplacian of the magnetic field
would be simply radiated away in electromagnetic waves. In the
quasi-static limit, that is, in the limit in which the timescale
associated with magnetic-field variations is large compared to the
light travel time $L/c$ (where $L$ is the typical lenghtscale of field
variations), any episodic time variation in the magnetic field
produced by the dynamics in the stellar interior would be rapidly
radiated away, leading to an exterior magnetic field which is again
with $\Delta \vec{B}\sim 0$. In our system, where the quasi-static
limit represents a good approximation, a similar behaviour outside the
star can be obtained by evolving the magnetic field according to a
diffusion equation\footnote{Note that the same equation holds for the
  vector potential, $\partial_t \vec{A} =\eta_0 \Delta \vec{A}$.},
$\partial_t \vec{B} =\eta_0 \Delta \vec{B}$. In this case, the
non-zero Laplacian components of $B$ are removed not through the
propagation of electromagnetic waves, but through resistive
dissipation. Hence, as long as one is not interested in the precise
dynamics of the exterior magnetic field, the two recipes are
effectively equivalent and the addition of a resistive term in
eq.~(\ref{a}) has the effect of mimicking the behaviour of the Maxwell
equations in the vacuum outside the star, at least within the
quasi-static approximation. Indeed, this approach is not novel, but it
has already been considered in the literature, for instance, by
\citet{BraithNord2006}.

\begin{figure*}
  \begin{center}
     \includegraphics[angle=0,width=5.4cm]{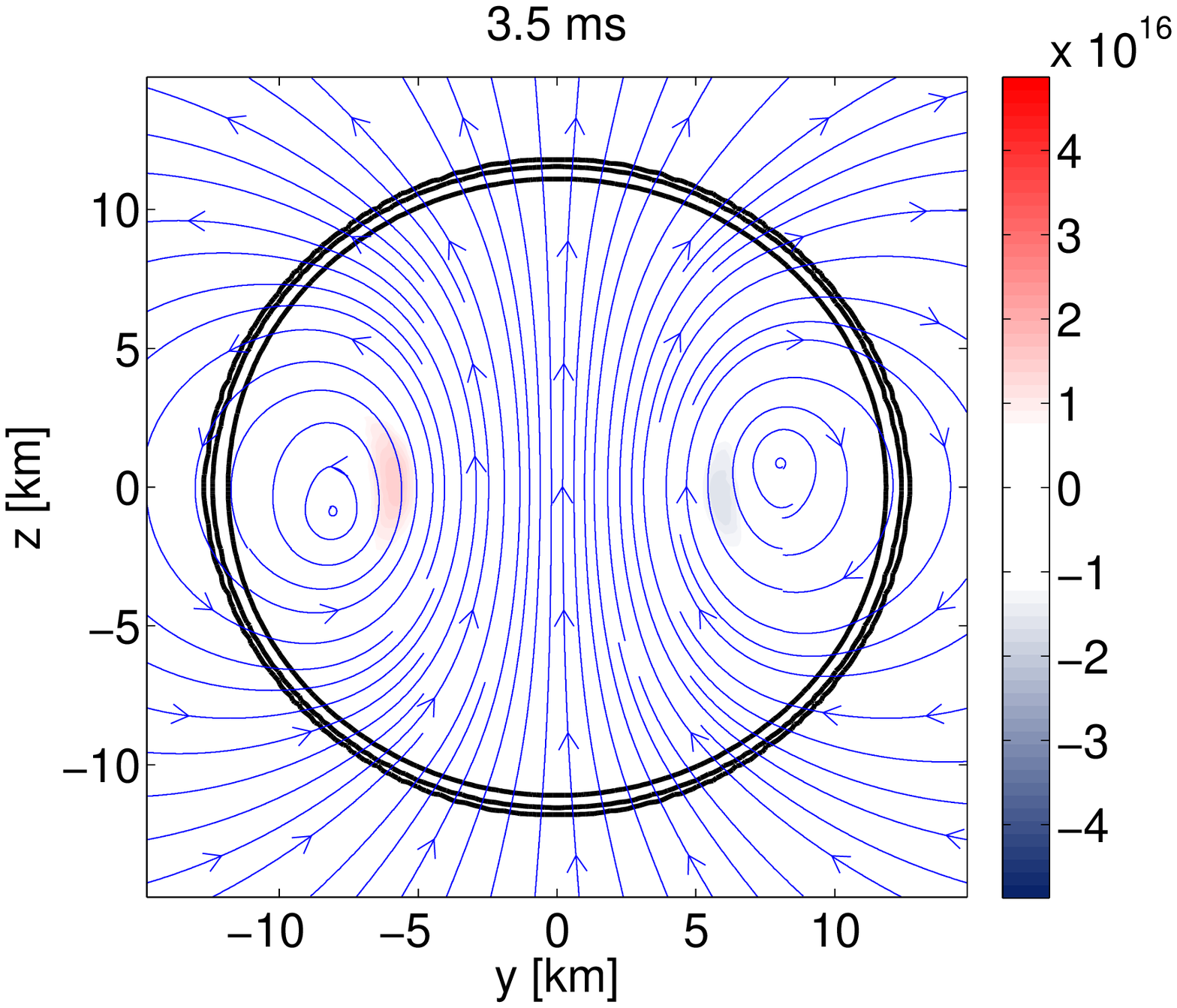}
     \hskip 0.1cm
     \includegraphics[angle=0,width=5.4cm]{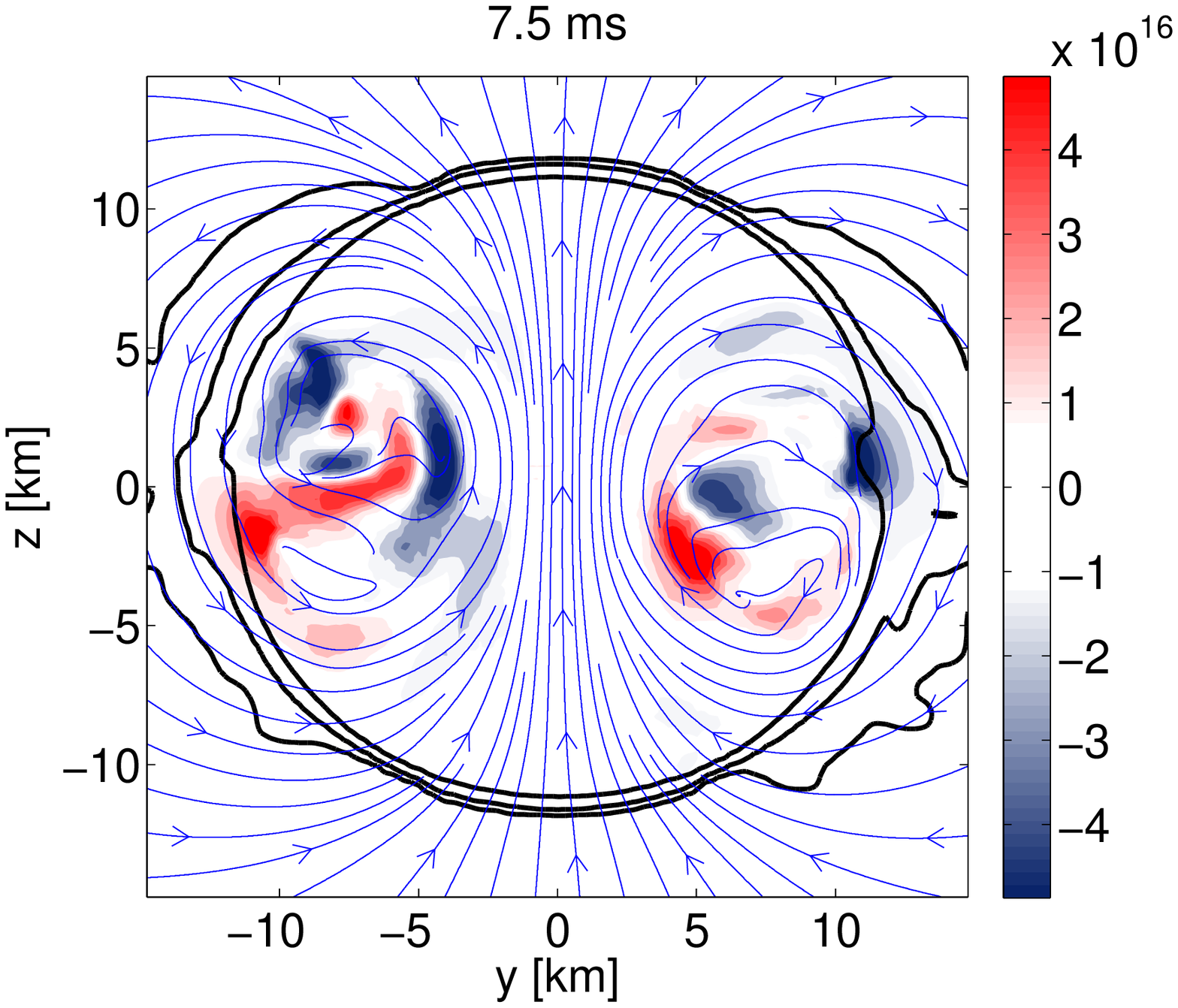}
     \hskip 0.1cm
     \includegraphics[angle=0,width=5.4cm]{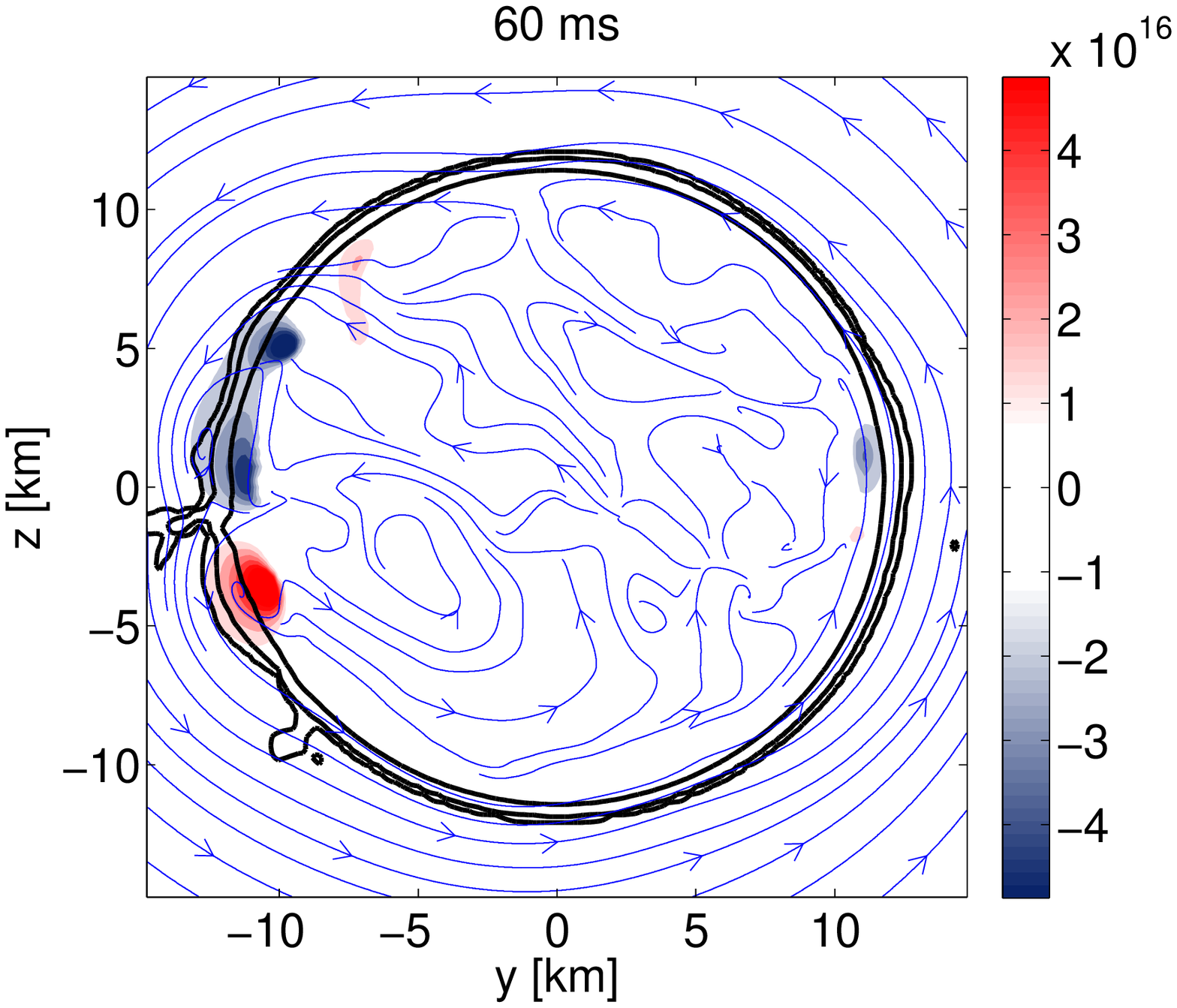}
     \includegraphics[angle=0,width=5.4cm]{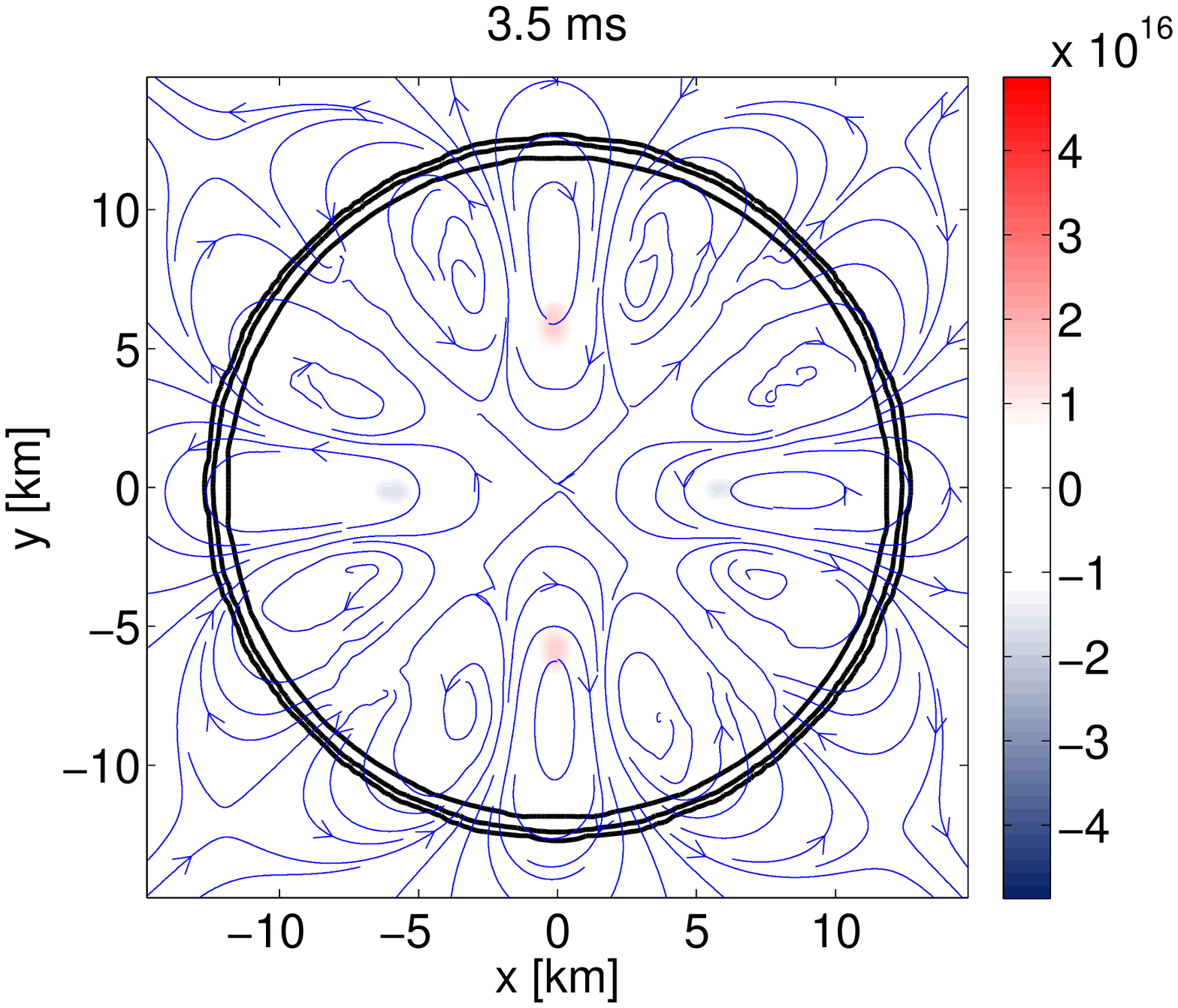}
     \hskip 0.1cm
     \includegraphics[angle=0,width=5.4cm]{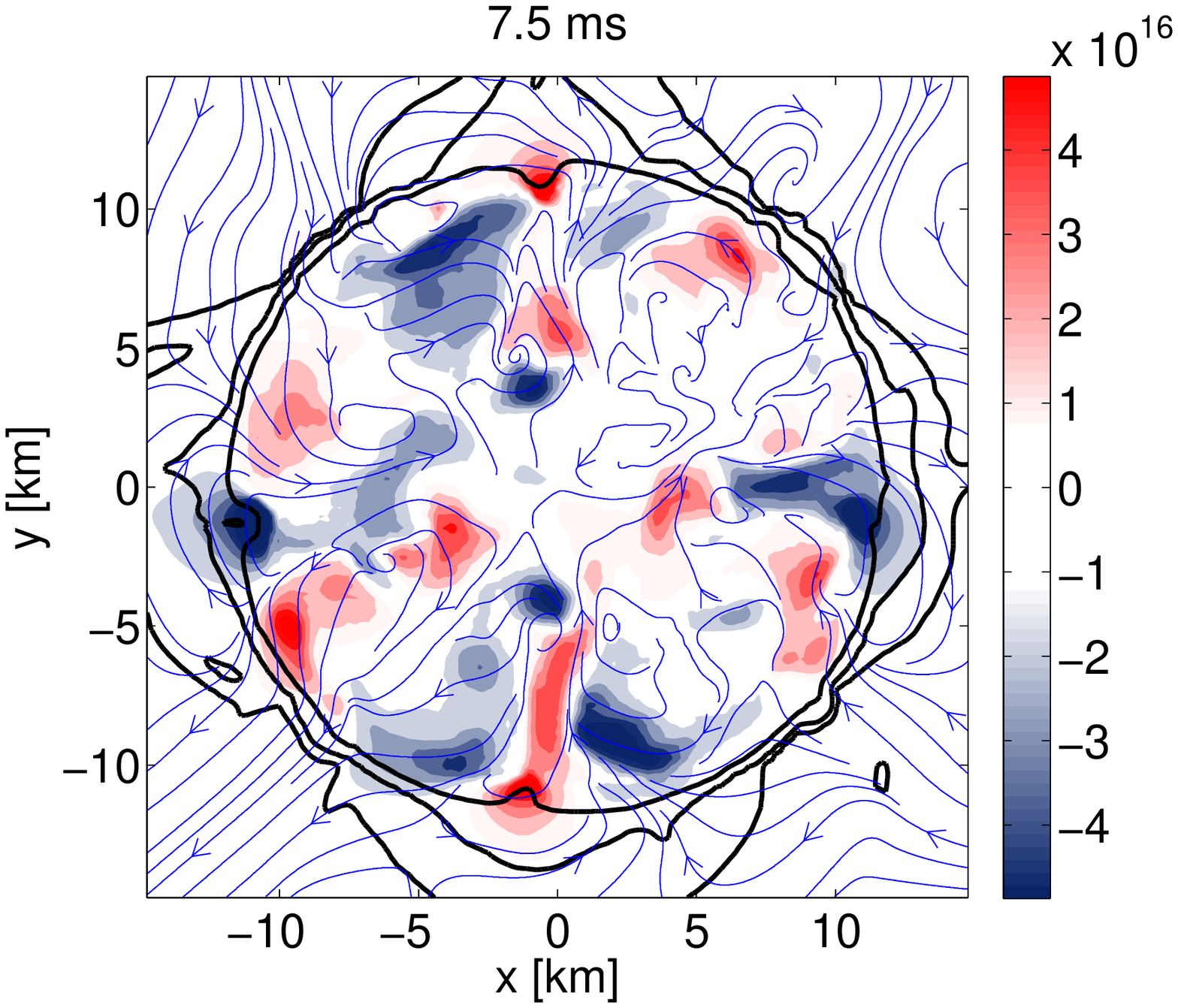}
     \hskip 0.1cm
     \includegraphics[angle=0,width=5.4cm]{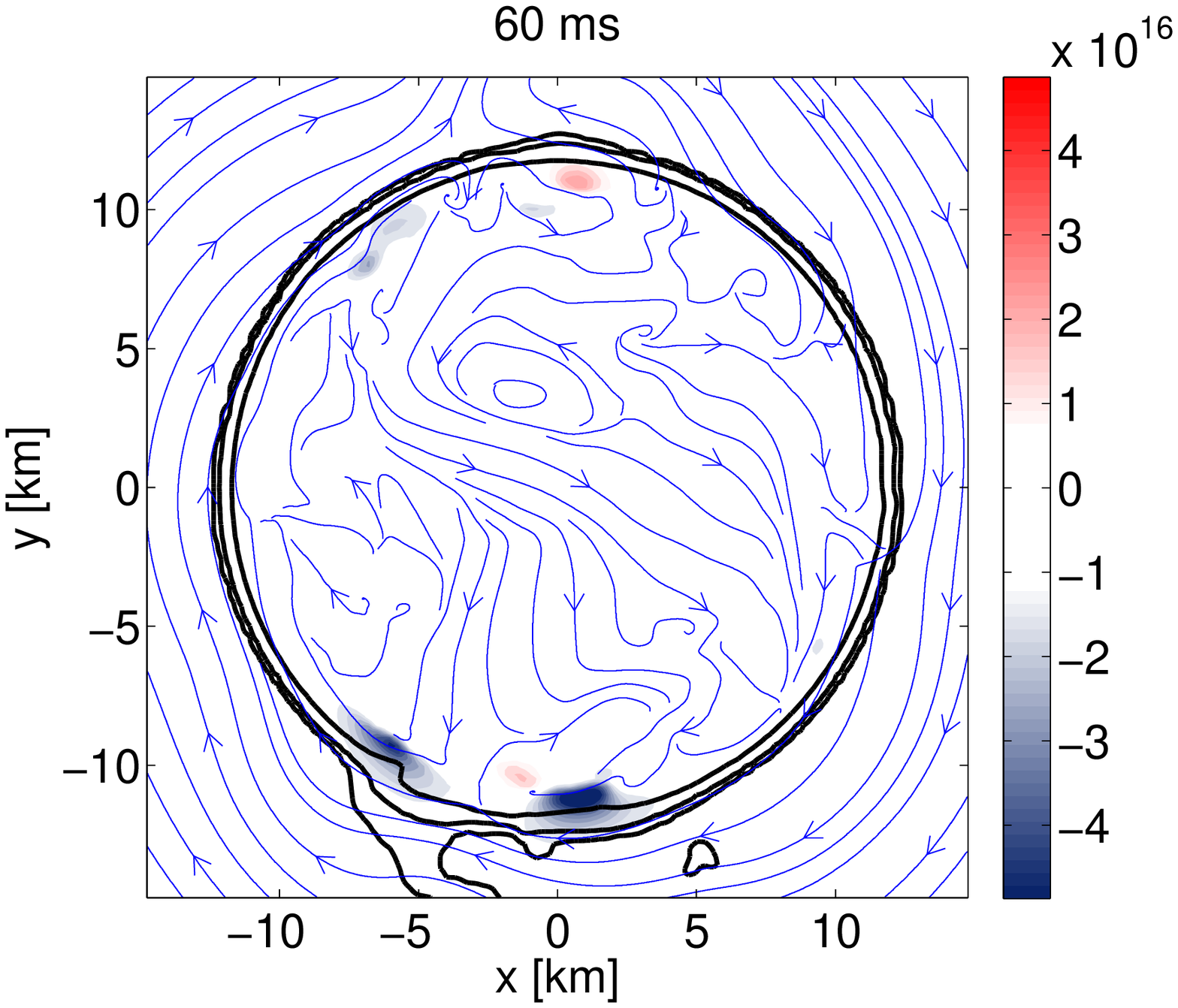}
  \end{center} 
  \caption{Poloidal field instability in our fiducial simulation with
    $B_{\rm p}=B_{6.5}$. Panels show different stages of the
    evolution (from left to right t = 3.5, 7.5, 60 ms) with meridional
    view (top row) and equatorial view (bottom row). Shown with
    vector lines are the (global) magnetic-field lines, while the
    colors show the intensity (in Gauss) of the toroidal magnetic
    field only; also reported are three rest-mass isodensity contours
    near the stellar surface, corresponding to $\rho=(0.02, 0.2, 2)
    \times 10^{13}\,{\rm g/cm}^3$. }
\label{fig:2Dplots}
\end{figure*}

This logic is valid as long as the external evolution takes place in
the quasi-static limit, and we fulfil this requirement by suitably
choosing the timescale of magnetic diffusion via the choice of the
resistivity $\eta_0$. More precisely, we set $\eta_0$ to be always
high enough (hence with sufficiently short associated resistive
timescales) so as to keep the exterior of the star always close to the
condition $\Delta \vec{B}\sim 0$. Although only an approximation, the
approach discussed above allows us to have a dynamical exterior
magnetic field, which adjusts itself to the changes triggered in the
interior by the development of the instability. The price to pay is
that, in addition to the one lost because of the numerical
resistivity, the magnetic energy of the system is not
conserved. Per-se, this would not be particularly problematic, since in a
realistic system one would expect the magnetic energy to be radiated
away, but it does introduce a dependence of the problem on the profile
chosen for the resistivity, since different profiles will be
responsible for energy losses.

As we will discuss in more detail the following Section, in fact,
during the initial development of the instability, the magnetic-field
modifications are still very small compared to the background field,
and the system is not expected to show visible changes, with the
magnetic energy being essentially conserved. Yet, because of the
resistive layer inside the star introduced via eq.~\eqref{a},
magnetic-energy losses will be present even in the early stages of the
evolution, simply because the magnetic field will be divergence free
but not Laplacian-free in the outer layer of the star.

In our previous work \citep{Ciolfi2011}, we had chosen a rather wide
resistive transition layer (\cf red dashed line in
Fig.~\ref{fig:diffusion}), which resulted in significant
magnetic-energy losses from the very beginning of the simulation. To
minimize these losses, we have here considered a thinner resistive
transition layer (\cf blue solid line in Fig.~\ref{fig:diffusion}),
thus involving a smaller stellar volume of the star where resistivity
can act. As a result, the magnetic-energy losses during the
exponential-growth phase (\ie during the first $\sim 3.5$ ms) go from
$\sim 30\%$ in the case of a wide layer, to below $\sim 2\%$ in the
case of a thin layer. In the subsequent stages of the instability, the
differences between the two prescriptions are much smaller and this is
because the dynamics of the field is much less influenced by the
properties of the resistive layer at the surface of the star. Yet,
since we have shown that different layers lead to different energy
losses, it is reasonable to wonder whether the properties of the
transition region can be important also for the subsequent evolution
of the system. As we will argue in detail in the Appendix, results are
effectively independent of $\eta_0$ as long as a suitable value for
$\eta_0$ is chosen for the different magnetic field strengths; more
specifically, if the resistivity is chosen to scale linearly with
$B_{\rm p}$ as $\eta_0/M_\odot=\eta_0^* \times(B_{\rm p}/B_{6.5}) = 0.1\times(B_{\rm p}/B_{6.5})$
(see Appendix).

\begin{figure*}
  \begin{center}
     \includegraphics[angle=0,width=5.9cm]{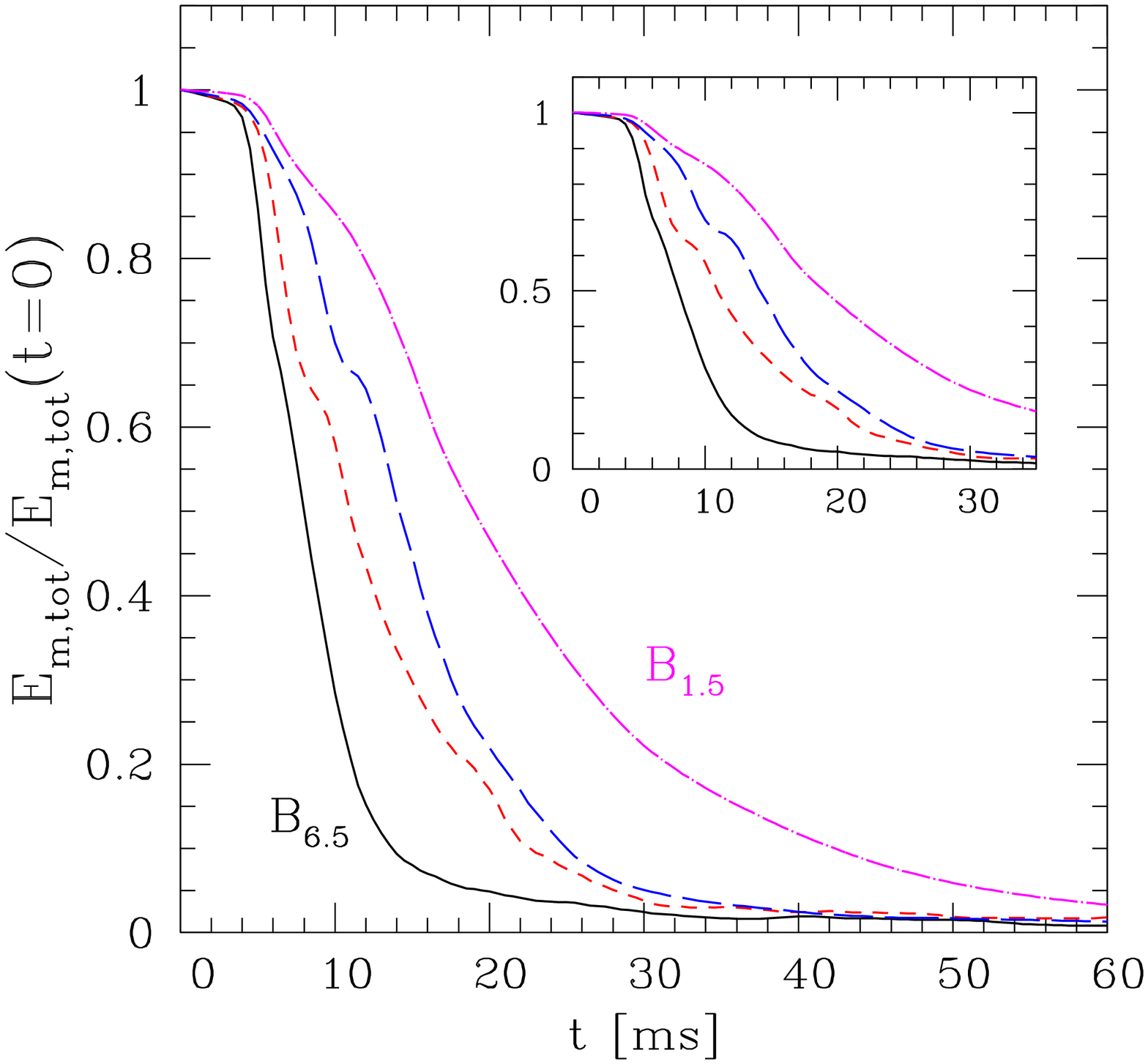}
     \hskip 0.1cm
     \includegraphics[angle=0,width=5.9cm]{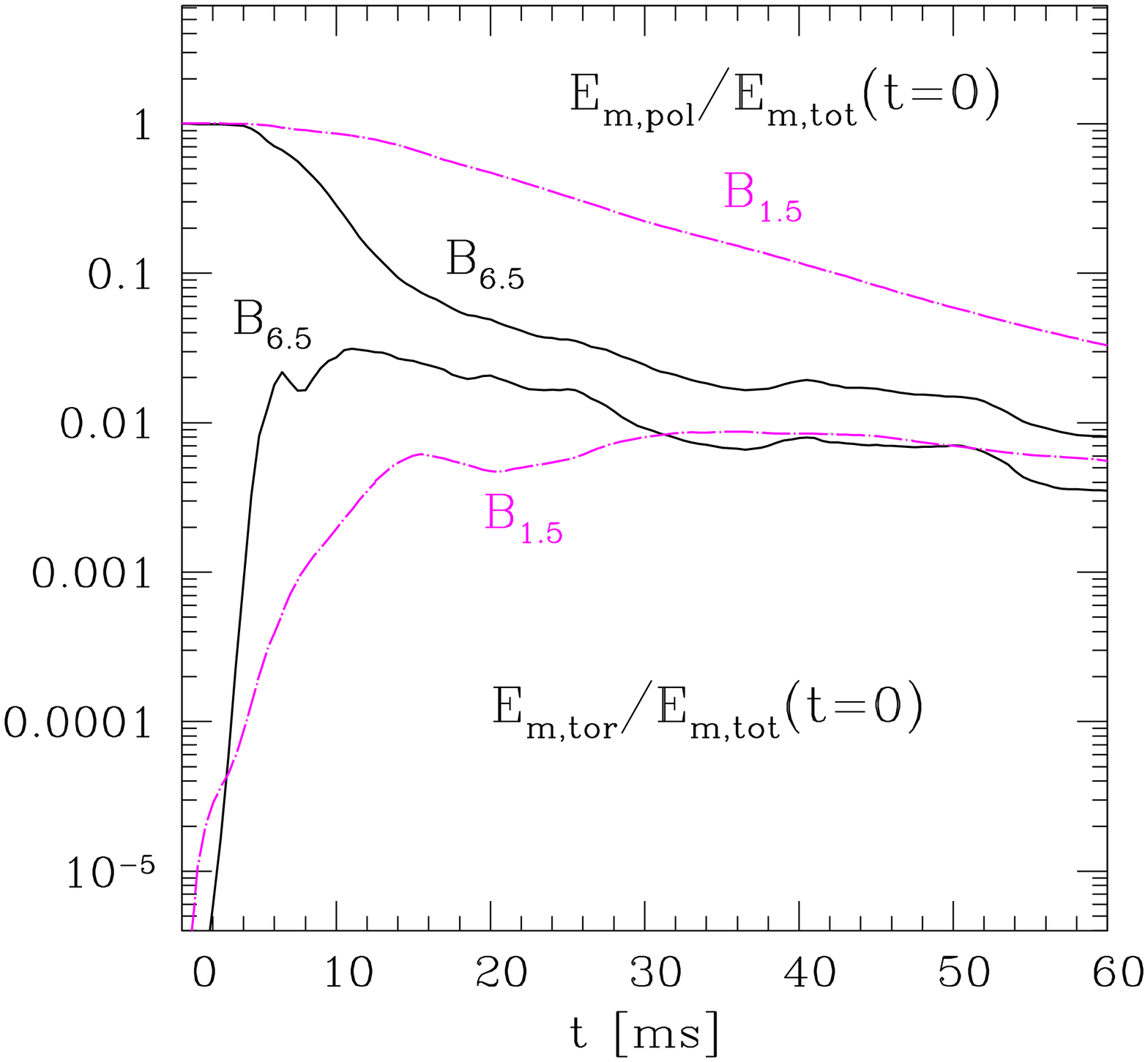}
     \hskip 0.1cm
     \includegraphics[angle=0,width=5.9cm]{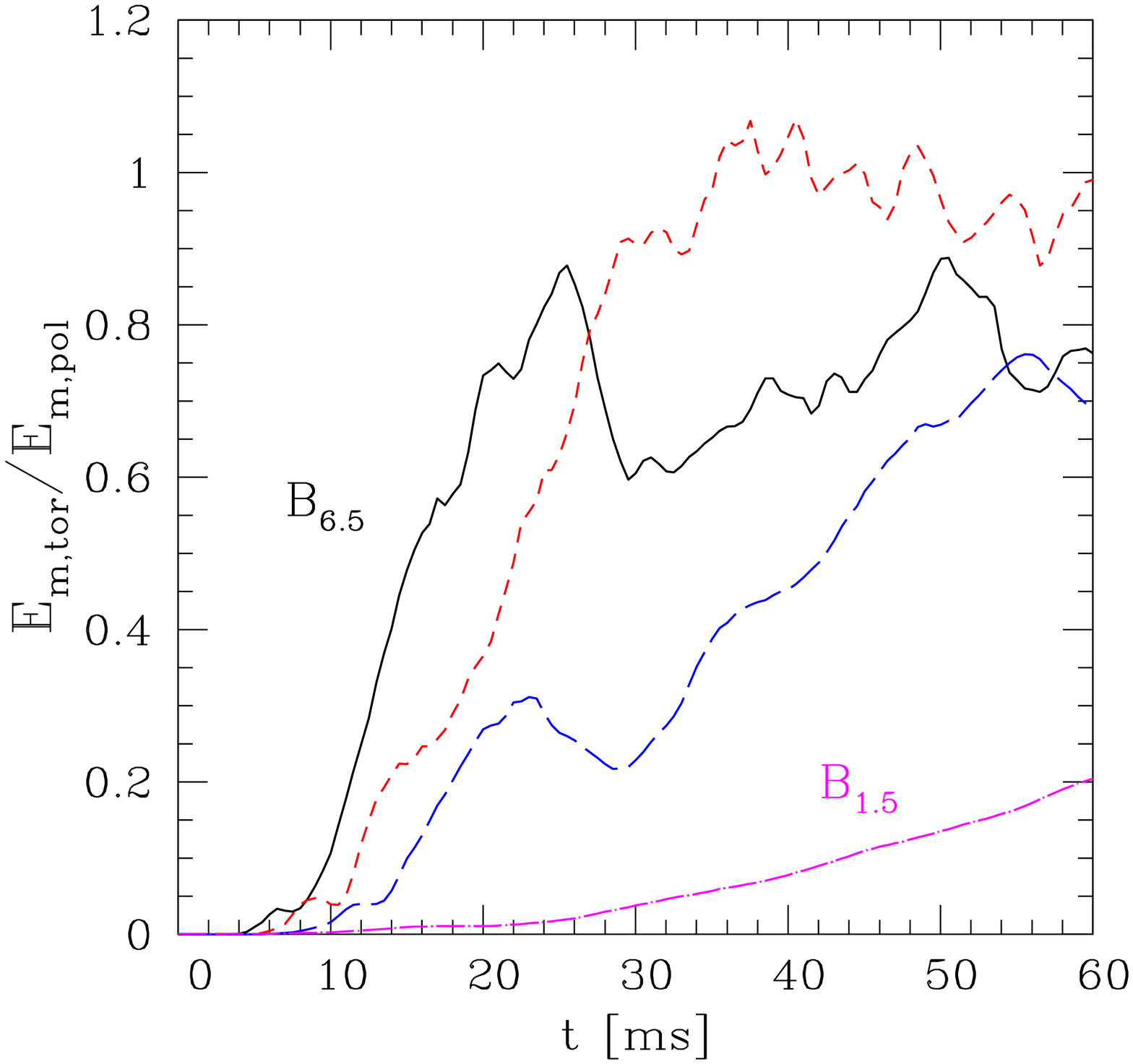}
  \end{center} 
  \caption{ {\it Left Panel}: Evolution of total magnetic energy
    normalized to the initial value, for different initial magnetic
    field strengths: $B_{6.5}$ (continuous black line), $B_{5.0}$ (dashed
    red line), $B_{3.5}$ (long-dashed blue line) and $B_{1.5}$ (dot-dashed
    magenta line). {\it Middle
      panel}: Evolution of poloidal and toroidal magnetic energies
    normalized to the initial total magnetic energy (log scale), for
    $B_{6.5},B_{1.5}$. {\it Right panel}: Ratio of toroidal and
    poloidal magnetic energies versus time, for the same collection of
    simulations shown in the left panel.
  }
\label{fig:energies}
\end{figure*}


\section{Results}\label{results}

\subsection{Poloidal field instability}\label{evolution}

We next describe the evolution of the system for our fiducial set of
initial models, from the onset of the instability to the nonlinear
rearrangement of magnetic fields. First of all, we briefly recall the
basic expectations of the perturbative studies on 
the onset of the instability for a purely poloidal field in nonrotating 
magnetized stars \citep{Markey1973,Wright1973}: (i) the instability first
develops in the region of closed-field lines surrounding the neutral
line; (ii) toroidal magnetic fields are generated in this region and
grow exponentially until their local intensity is comparable to the
poloidal one, which corresponds to the saturation of the instability;
(iii) the instability saturates in about one Alfv\'en time\footnote{An
  estimate of the Alfv\'en time is given by $\tau_A \sim
  2R\sqrt{4\pi\langle\rho\rangle}/B_{\rm p}$, where $\langle\rho\rangle$
  is the average rest-mass density. Hence, $\tau_A\sim 3$ ms for our
  fiducial model with $B_{\rm p}=B_{6.5}$.}; (iv) the timescale
associated with the exponential growth of the toroidal field scales as
$1/B$ with the magnetic field strength. As we will show in the
following, our simulations meet all of these expectations.

A first visual overview of the dynamics of the system is given in
Fig. \ref{fig:2Dplots}, where we present snapshots of the star and its
magnetic field in the meridional $(y,z)$ (top row) and equatorial
$(x,y)$ planes (bottom row) at three different stages of the
evolution. In this example $B_{\rm p}=B_{6.5}$. The left panels refer
to the time of saturation of the instability, when the magnetic field
starts to show visible modifications ($t=3.5$ ms). From the equatorial
view we note that the initial axisymmetric geometry is lost and
replaced by a non-axisymmetric structure. Around 7.5 ms (central
panels) the instability has fully developed. As expected, the closed
line region is filled with toroidal fields of strength comparable to
the background, resulting in vortex-like structures of magnetic field
lines. This stage is the most dynamical and the rapid modifications of
the field lead to the expulsion of matter from the surface of the star
($\sim 2\times10^{-4}\,M_\odot$ in this example). Up to this point in
the evolution the instability has affected only the region of closed
magnetic field lines, in accordance with the perturbative predictions.
However, as the nonlinear rearrangement of the magnetic field
proceeds, the whole star is involved, with changes encompassing also
the open field lines and the exterior field. Of course, the extent of
these modifications will depend on the strength of the magnetic field,
being larger for more violent field dynamics. The last panels refer to
the end state of our simulation ($t=60$~ms). At this stage there is no
trace of the initial geometry and the system has lost most of its
magnetic energy (see discussion below). To make sure that the external
magnetic field is always close to a potential one (even if its
geometry is unusual, as in this case) we have computed its curl and
found it is always smaller than $\sim 10^{-6}$, while inside the star
and near the surface it reaches $\sim 10^{-3}$. The same result holds
for all the cases considered in this work\footnote{An essentially
  potential external magnetic field is not surprising given that the
  field in the stellar exterior is to a good precision with zero
  divergence and zero Laplacian.}. In comparison with the snapshots
presented in \citet{Ciolfi2011} for the same magnetic field strength,
the evolution appears more violent. The significant magnetic energy
loss due to resistivity in a thicker transition layer, in fact, had
the effect of restraining the dynamics, resulting in a less dramatic
and more ordered evolution.

\begin{figure*}
  \begin{center}
     \includegraphics[angle=0,width=5.4cm]{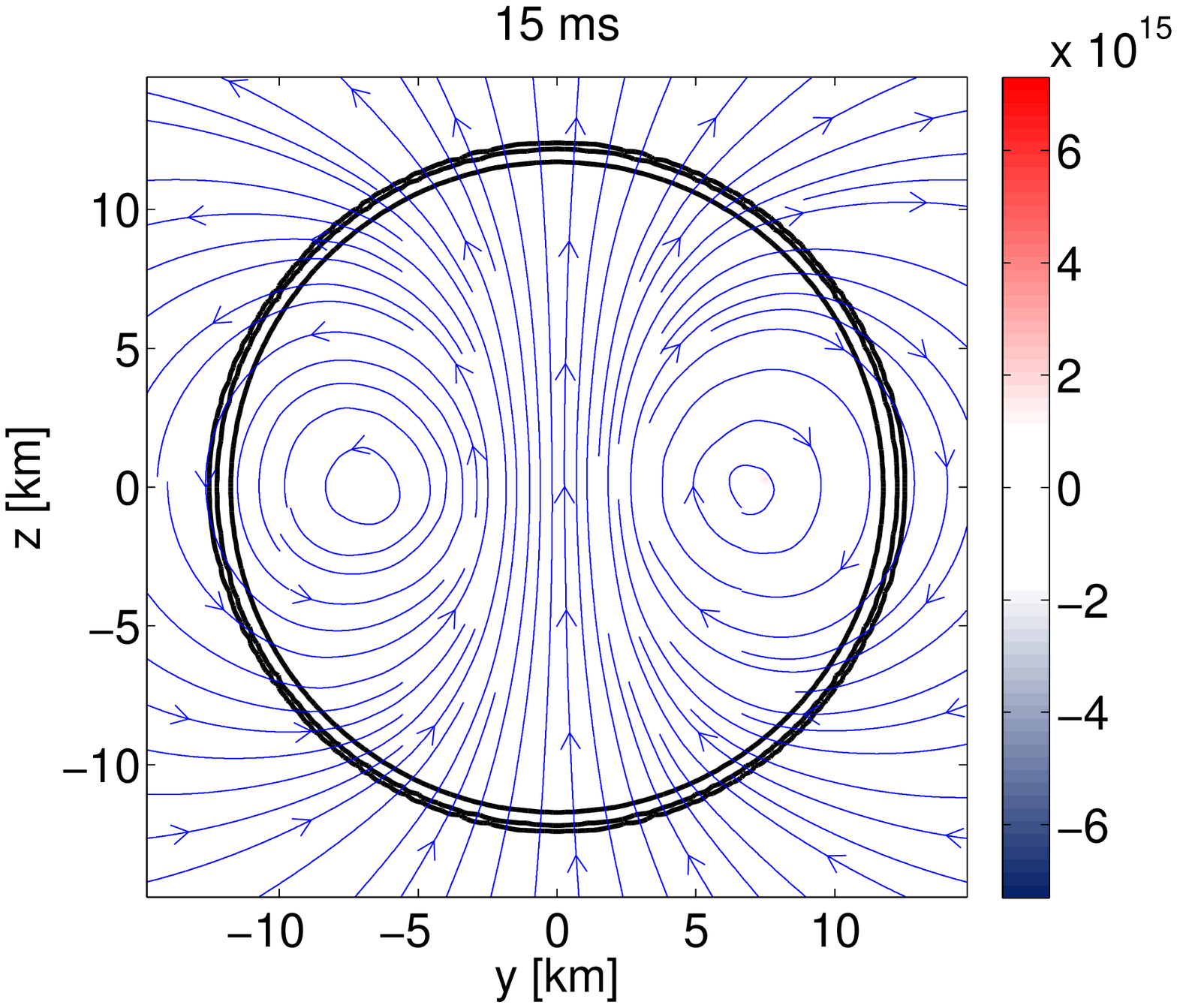}
     \hskip 0.1cm
     \includegraphics[angle=0,width=5.4cm]{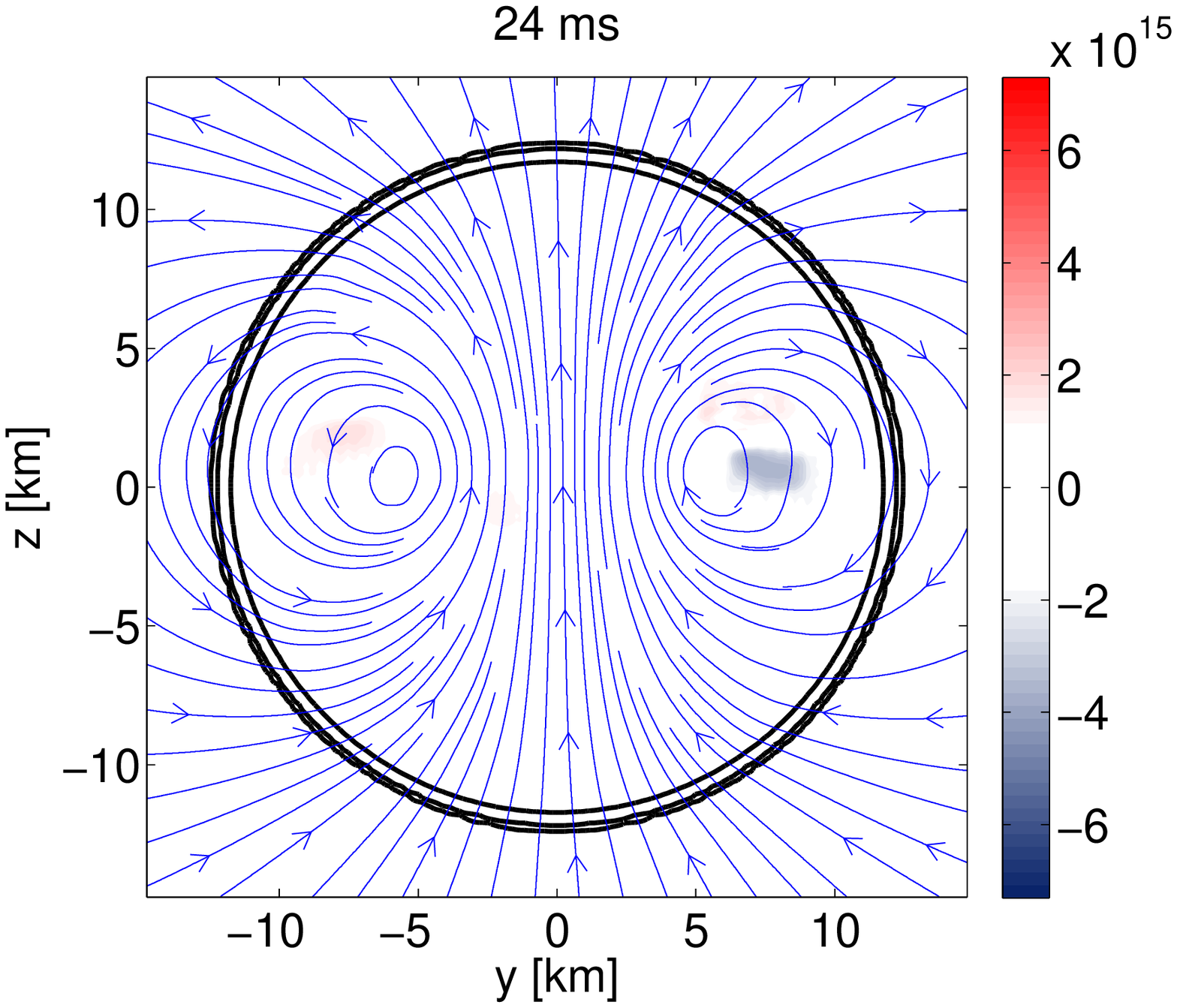}
     \hskip 0.1cm
     \includegraphics[angle=0,width=5.4cm]{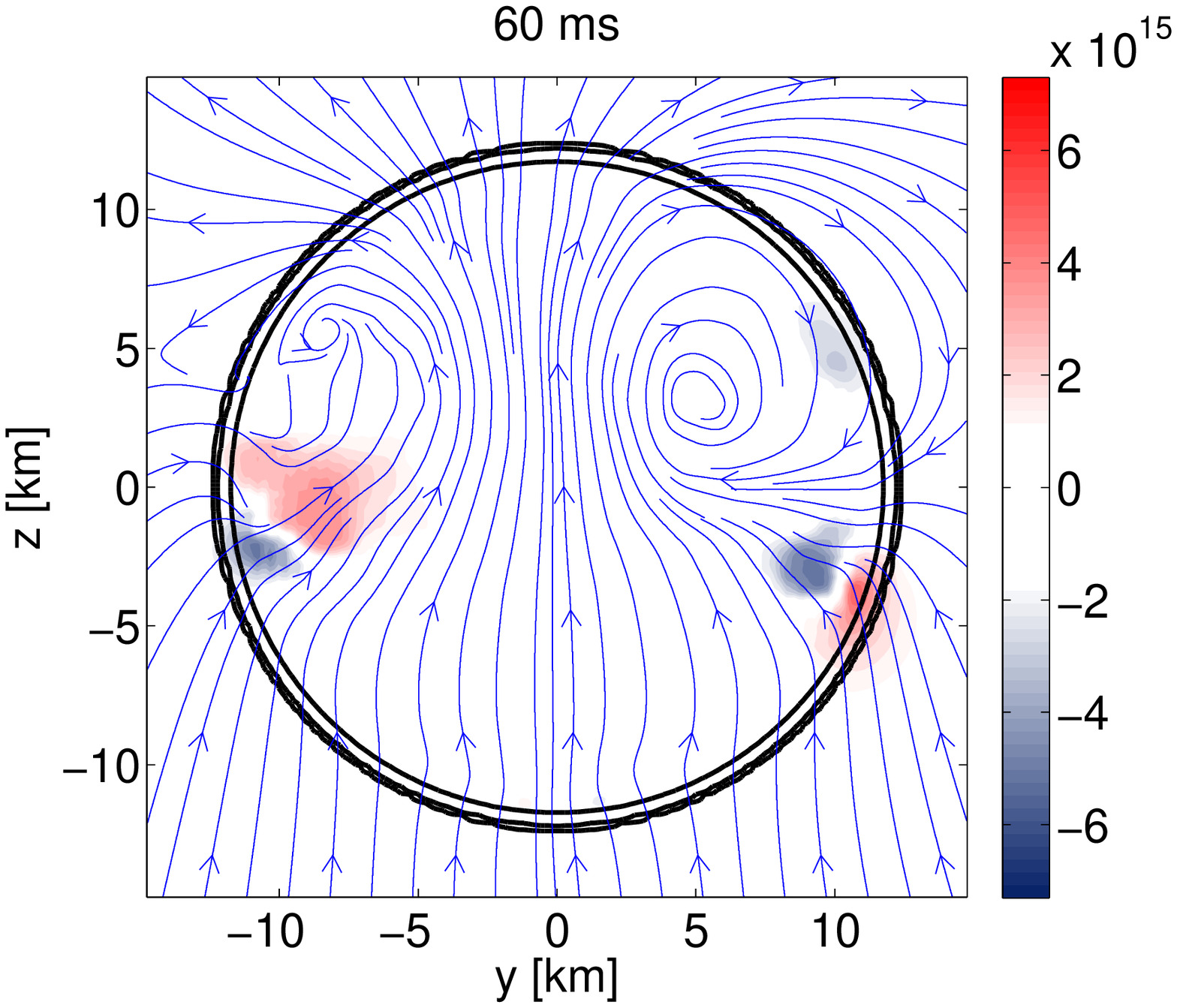}
     \includegraphics[angle=0,width=5.4cm]{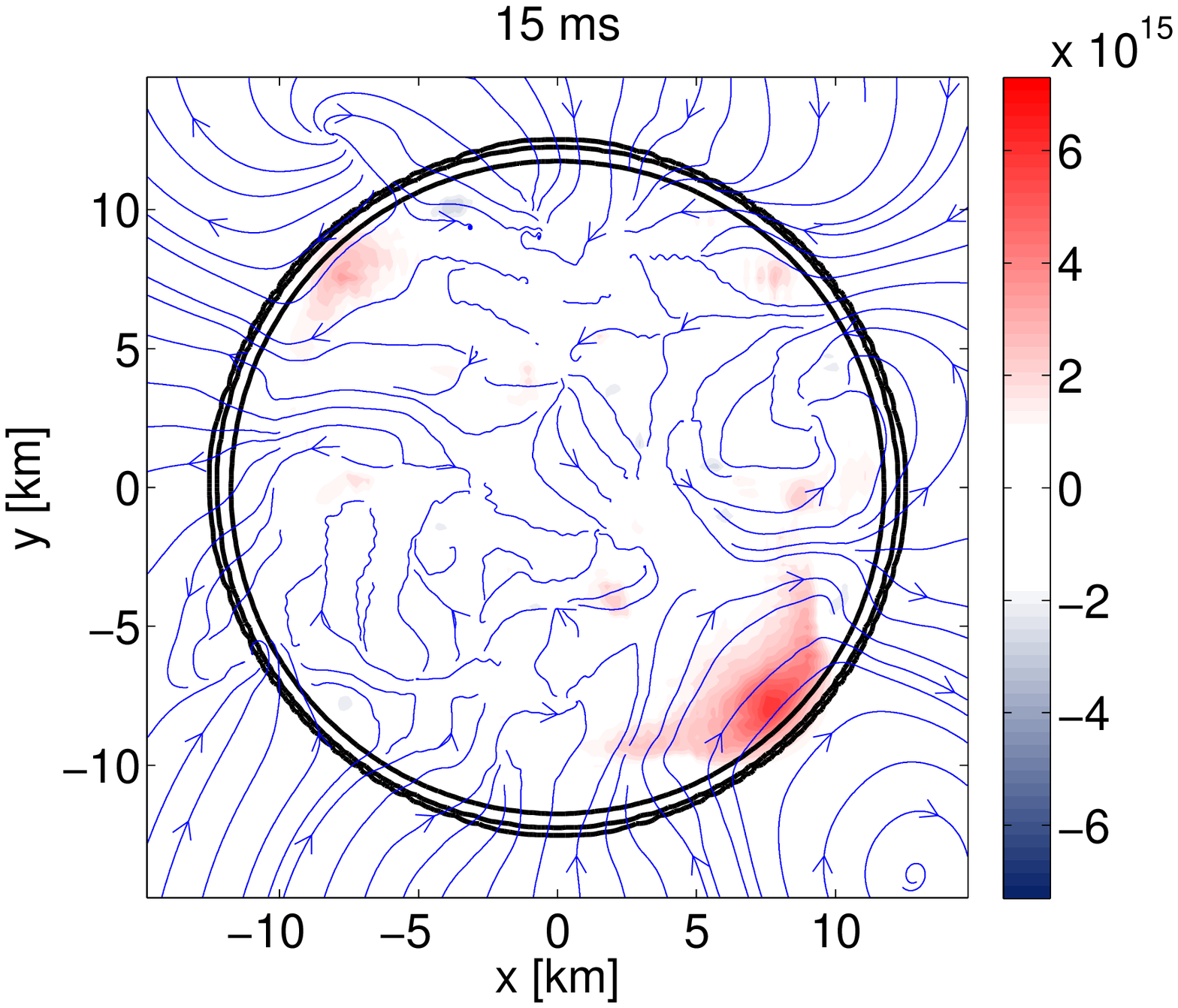}
     \hskip 0.1cm
     \includegraphics[angle=0,width=5.4cm]{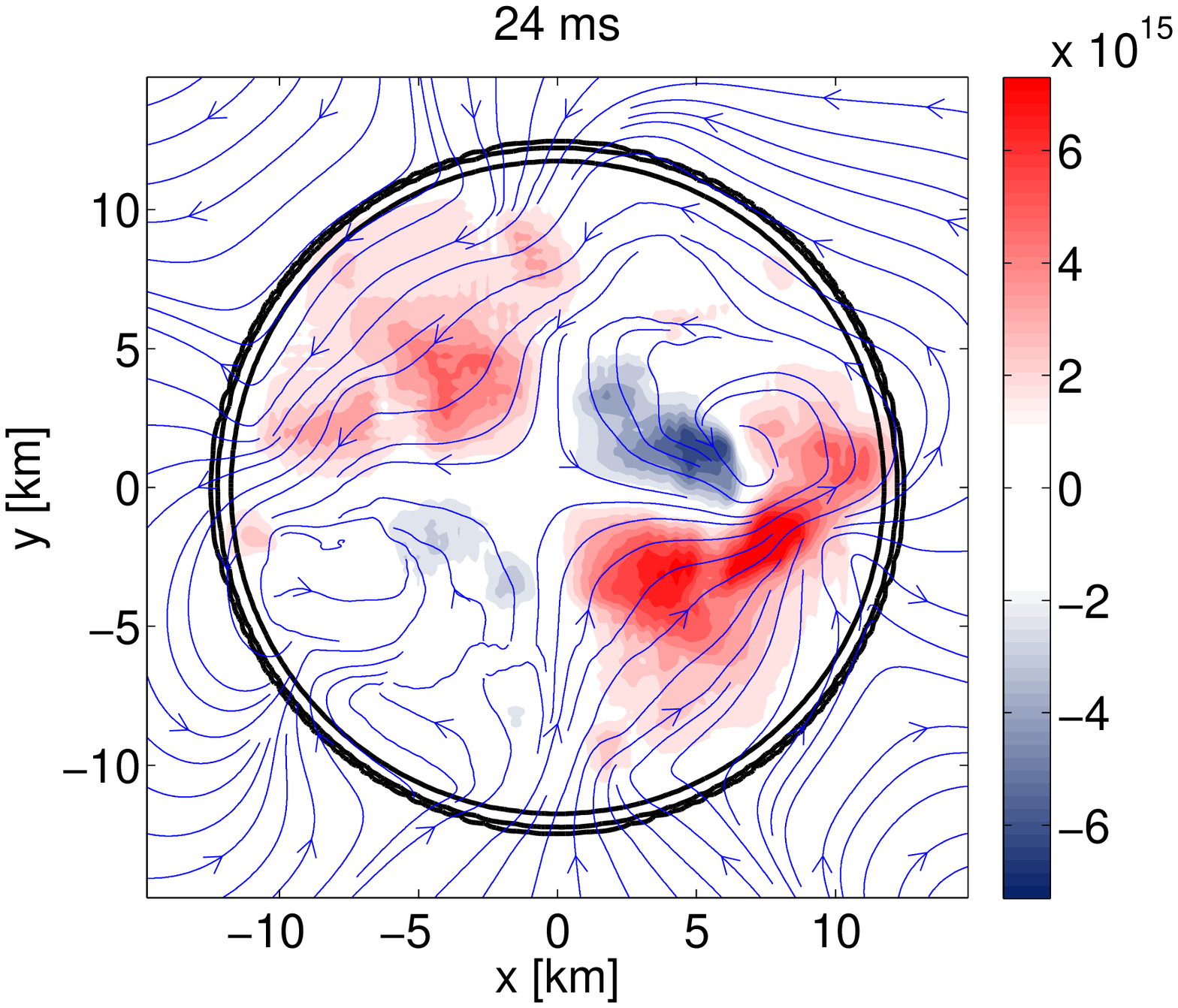}
     \hskip 0.1cm
     \includegraphics[angle=0,width=5.4cm]{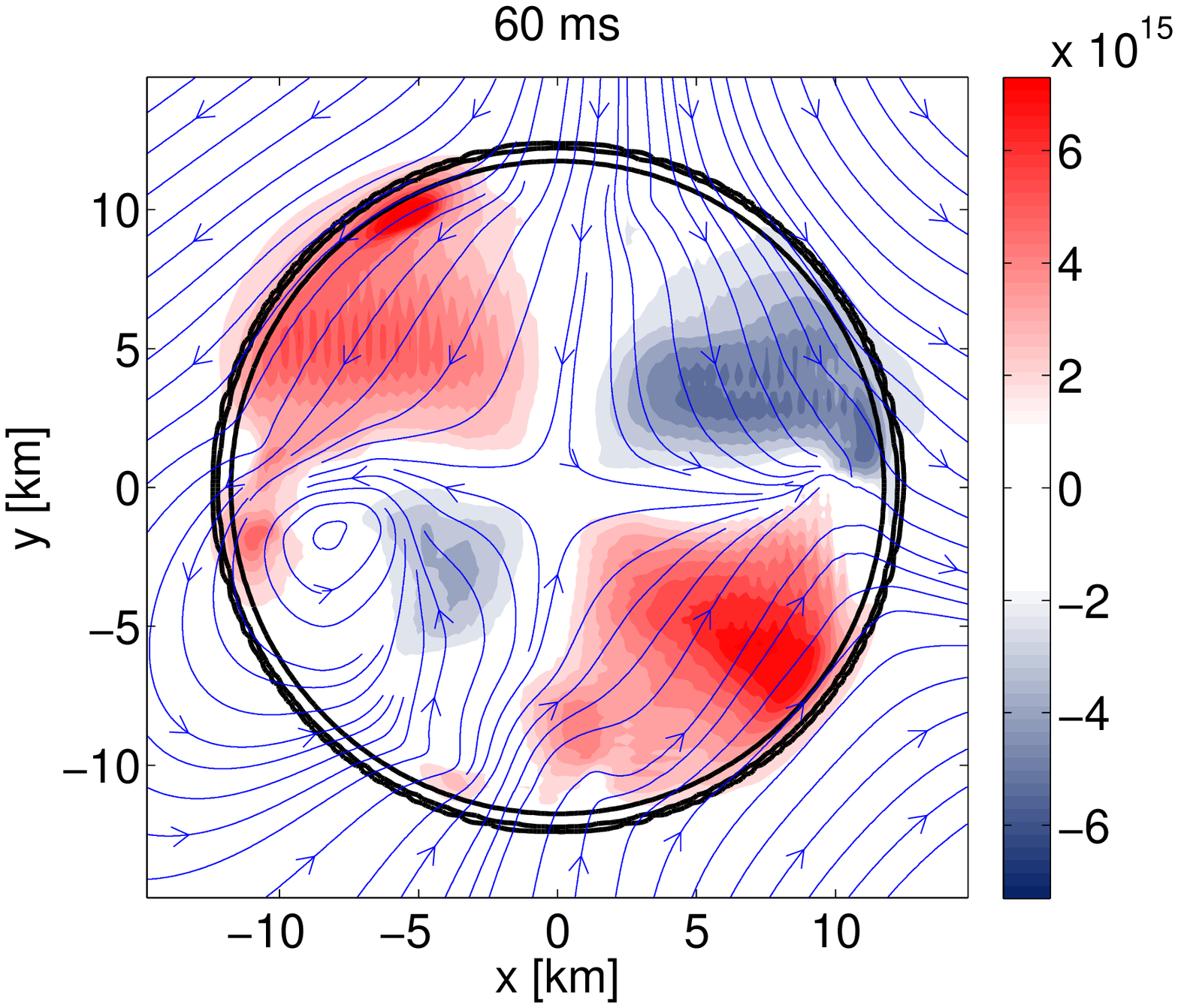}
  \end{center} 
  \caption{The same as in Fig.~\ref{fig:2Dplots}, but for a star with
    initial magnetic field of $B_{\rm p}=B_{1.5}$. Note that the
    dynamics is less violent and that the color scale is different
    from the one employed in Fig.~\ref{fig:2Dplots}.}
  \label{fig:1.5_2D}
\end{figure*}

In Fig.~\ref{fig:energies} we plot the evolution of the poloidal,
toroidal and total magnetic energies. In practice, the evolution of
the total magnetic energy (left panel) provides information on the
various stages of the evolution: the initial phase up accompanied by
very small variations, the sudden drop of magnetic energy with a loss
of $\sim 90$\% of magnetic energy in few ms to tens of ms depending on
the initial magnetic field strength, and the final stage with a slower
evolution as the energy decreases to few percent of the initial value
and the system reaches a quasi-stationary equilibrium\footnote{Note
  that we refer to a ``quasi-stationary equilibrium'' and not to a
  stable-equilibrium because, strictly speaking, the latter is
  impossible to prove with numerical simulations over a finite amount
  of time.}. Assessing the equilibrium properties of the new
configuration is complicated by the fact that, by construction, our
system would suffer from resistive losses due to a thin layer inside
the star and outside of it, even if a stable equilibrium had been
reached. Hence, these residual losses make it difficult to determine
unambiguously whether the new, post-instability configuration is a
stationary one or not. What is evident is that the system's properties
are not changing significantly (\eg the rest-mass density) or, if
changing, they are doing so much more slowly than during the 
instability (\eg the magnetic field). We interpret this behaviour as (partial)
evidence that the new magnetic-field configuration has reached a
quasi-stationary state or that, if still intrinsically unstable, the
growth time of the instability is much larger than the one we can
possibly investigate. As we will comment below, this conclusion is
further supported by the dynamics of the rest-mass density.

The overall evolution does not change with smaller initial field
strengths, apart from the expected increase of
timescales. Figure~\ref{fig:energies} also reports the poloidal and
toroidal field energies in a logarithmic scale (middle panel). Note
that the the toroidal energy grows exponentially in the initial stages
of the evolution and that the growth rate depends on the initial
magnetic-field strength, the slope being steeper for larger magnetic
fields. The poloidal component, on the other hand, remains almost
unchanged during this initial stage, until the nonlinear
rearrangement of the magnetic field starts. Then, the system evolves
losing most of the poloidal energy, while the toroidal energy
experiences smaller variations, thus resulting in a growing ratio of
toroidal and poloidal energies. In the last phase of the evolution,
the ratio of the poloidal-to-toroidal magnetic energies stabilizes in
the range $\sim 0.7-1$, indicating that at this stage the toroidal
magnetic field provides a dynamically important contribution to the
final quasi-stationary balance (right panel of Fig.~\ref{fig:energies}).

From a closer look at the details of the dynamics, we notice a
qualitative difference between stronger and weaker magnetic fields,
the latter having a much smoother evolution. For $B_{\rm p} =
B_{1.0}-B_{2.5}$ the system undergoes indeed less dramatic
modifications, where the initial magnetic field geometry is partially
maintained\footnote{In particular, the open field lines and the exterior 
field are not strongly affected as in the case of stronger fields 
(see Fig.~\ref{fig:2Dplots}). This result is more similar to what obtained 
in \cite{Lasky2011}.}. This is evident in Fig.~\ref{fig:1.5_2D}, where we 
show snapshots of the evolution of a star with initial magnetic field 
strength $B_{\rm p} = B_{1.5}$. These differences are relevant when we 
consider the emission properties of the system and, in particular, the 
emission of gravitational waves.
 
\begin{figure}
  \begin{center}
     \includegraphics[angle=0,width=8.0cm]{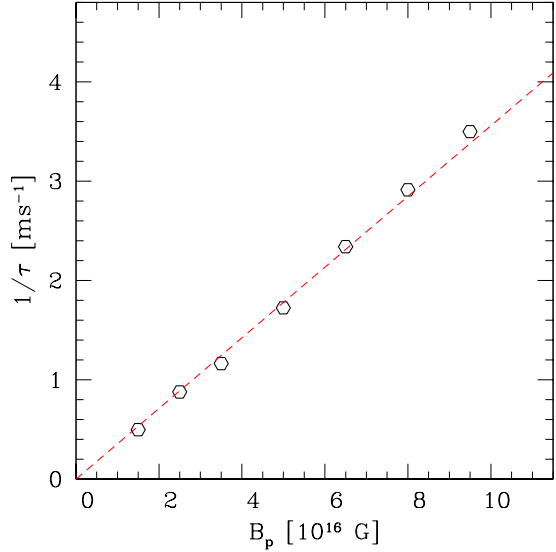}
  \end{center} 
  \caption{Instability growth time as a function of the initial
    magnetic field strength (black circles). The red dashed line
    represents a fitted linear scaling as expected from perturbation
    theory.}
\label{fig:tauscaling}
\end{figure}

In Fig.~\ref{fig:tauscaling} we show instead the growth time $\tau$ of
the toroidal field energy in the exponential phase of the evolution,
as a function of the initial magnetic field strength. The predicted
linear scaling is very well satisfied in the full range of magnetic field 
strengths considered (the red dashed line represents a linear fit to 
the data). 


\subsection{Final magnetic field configuration}\label{endstate}

Since the new simulations have been carried out on timescales which
are much longer than the ones investigated in \citet{Ciolfi2011}, we
are in a much better position to discuss the properties of the final
magnetic-field configuration. Also important in this context is the
evolution of purely-hydrodynamical quantities, which can provide
important additional information on the development of the instability
and on the quasi-stationary state approached by the system in the
final stages. One of such quantities is the central (\ie maximum)
rest-mass density. We recall that the initial purely poloidal magnetic
field is responsible for a quadrupolar deformation of the star, whose
resulting shape is slightly oblate (\ie with a positive ellipticity
$\epsilon$), despite being nonrotating. This deformation lowers the
central rest-mass density of the star with respect to the unmagnetized
case, and the difference is of the order of $\Delta
\rho_{max}/\rho_{max}(B=0) \equiv 1 - \rho_{max}/\rho_{max}(B=0)\sim
\epsilon$. Since the
ellipticity scales as $\epsilon \sim B^2$, when the system loses a
factor $\sim 10^2$ in poloidal magnetic energy, $\Delta\rho_{max}$
should be also reduced by about the same factor. In addition, if there
is a significant toroidal component, this tends to deform the star in
a prolate shape, contrasting the effect of the poloidal field. As a
consequence, in our simulations we expect the final $\Delta\rho_{max}$
to be much smaller than the initial one, \ie~the star basically
recovers the spherical shape of the unmagnetized case.

\begin{figure}
  \begin{center}
     \includegraphics[angle=0,width=8.0cm]{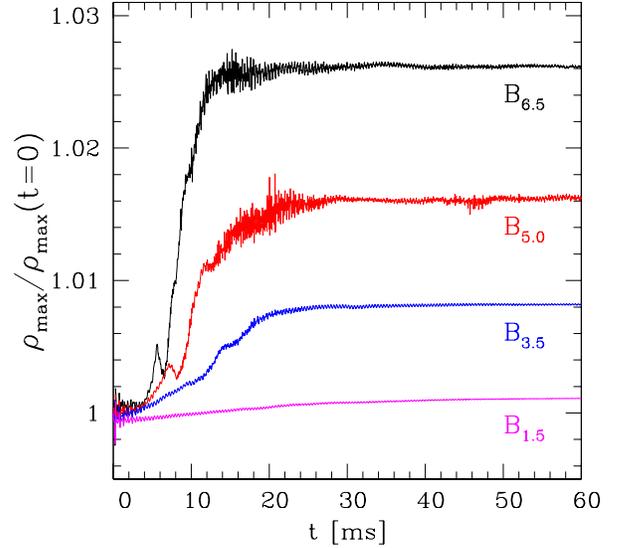}
  \end{center} 
  \caption{Evolution of central rest-mass density for different
    initial magnetic field strengths. From top: 6.5 (black), 5 (red),
    3.5 (blue), $1.5\times 10^{16}\,{\rm G}$ (magenta). }
\label{fig:rhomax}
\end{figure}

In Fig.~\ref{fig:rhomax} we show the evolution of the central
rest-mass density for different initial magnetic field strengths. Note
that our different models are all built with the same initial central
density, which gives a percent difference in the central density of
the unmagnetized star corresponding to each model. Clearly, also for
the rest-mass density the evolution is much more dramatic for
stronger initial magnetic fields, with excursions in the central
density that become larger and more rapid as the initial magnetic field strength
is increased. In all cases, however, the central density mimics the
evolution of the magnetic energy and reaches an approximately constant
value, with an overall jump which scales roughly as $\sim B^2$. The
absence of additional evolution in the rest-mass density after the
instability has fully developed provides the indication that the new
configuration is, at least hydrodynamically, stable, and that, if the
new magnetic-field configuration is about to lead to a new unstable
evolution, it will do so on much larger timescales. For initial
magnetic field strengths of $6.5$ and $5\times 10^{16}\,{\rm G}$, the change is
so violent and rapid that it leads to high-frequency oscillations in
the central density, while we do not observe the same effect with
smaller fields. As we will further discuss in
Sect.~\ref{gravitationalwaves}, a rapid variation of central density
can also affect the emission properties of the star, \eg introducing a
significant modulation of the gravitational-wave signal.

Other useful information on the final state can be obtained by looking
at the evolution of the total magnetic helicity $H_m$. This quantity
is associated with the topological properties of a given magnetic
field geometry, and measures the degree of linkage of magnetic field
lines \citep{Moffatt1969}. The initial purely poloidal configuration
has $H_m=0$ because only regions with a mixed poloidal-toroidal field
contribute to the magnetic helicity\footnote{We note, however, that even a
mixed field can be constructed such that $H_m=0$, \eg in axisymmetry,
through the superposition of a poloidal magnetic field and of two
toroidal ones of opposite polarity.}. We already noticed that the
ratio of toroidal and poloidal energies grows in time and then tends
to stabilize to values in the range $\sim 0.7-1$. The total magnetic 
helicity, even if related to the energy in the two magnetic-field 
components, encodes an independent information about
the topology of the field, and it is therefore useful to investigate
its evolution in our system.

We recall that although in ideal MHD the total magnetic helicity is
conserved \citep{Woltjer1958}, a highly-conducting fluid star 
in vacuum with a thin resistive layer does not represent an
ideal-MHD system, and magnetic helicity conservation is therefore not
guaranteed [of course helicity is also produced because of the
  intrinsic nonzero numerical resistivity, but the latter is much
  smaller than the one introduced in eq.~\eqref{a}]. We 
compute the total magnetic helicity as
\begin{equation}
H_m \equiv \int_{\Sigma_t} H_m^{0} \sqrt{-g}~d^3 x \,,
\label{Hm}
\end{equation}
where $H_m^{\alpha} \equiv {}^*F^{\alpha\beta}A_\beta$ is the magnetic
helicity 4-current and $\Sigma_t$ is the spatial hypersurface at a
given time $t$ [see \citet{Bekenstein} and references therein]. Since
our initial helicity is $H_m(t=0)=0$, we normalize $H_m$ to the
helicity of an axisymmetric twisted-torus configuration in which the
poloidal field is arranged as in our initial condition and the
toroidal field, confined in the region of closed field lines around the 
neutral line, is uniform and with an energy equal to the poloidal one,
 \ie \hbox{$E_{pol}=E_{tor}=E_m/2$}. Indicating with $\tilde{H}_m$ such
reference helicity, we can estimate it as $|\tilde{H}_m|\sim r_{_N} \,
\sqrt{E_{pol}~E_{tor}}=r_{_N}\, E_m/2$, where $r_{_N}$ ($\sim$8 km) is the radius
of the neutral line and $E_m$ is set to coincide with the magnetic
energy of the stellar model under exam. Note that we are not concerned
here with the sign of $H_m$ (or of $\tilde{H}_m$), since the latter
depends on whether toroidal fields are directed along the positive or
negative $\phi$-direction and a global transformation
$B^{\phi}\rightarrow -B^{\phi}$ would not change the properties of the
system.

\begin{figure}
  \begin{center}
     \includegraphics[angle=0,width=8.0cm]{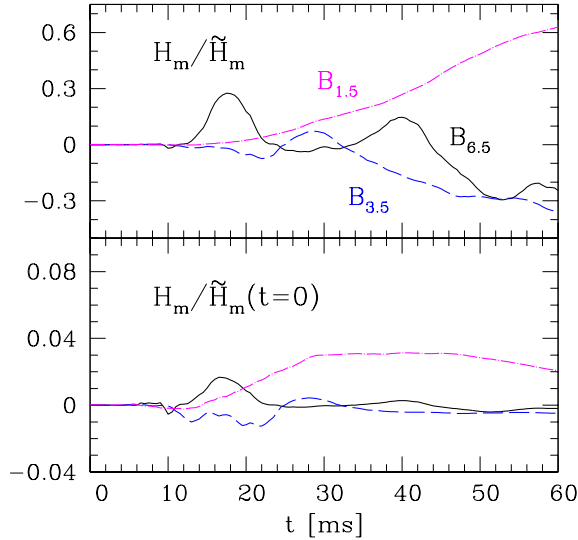}
  \end{center} 
  \caption{{\it Top}: Evolution of magnetic helicity normalized to
    $\tilde{H}_m$ (see text), for different initial magnetic field
    strengths: 6.5 (continuous black line), 3.5 (long-dashed blue
    line) and $1.5\times 10^{16}\,{\rm G}$ (dot-dashed magenta line).
    {\it Bottom}: Evolution of magnetic helicity normalized to the
    initial value of $\tilde{H}_m$, for the same set of simulations.}
\label{fig:NormHmag}
\end{figure}

We can now monitor the evolution of $H_m$. As long as 
$|H_m/\tilde{H}_m| \ll 1$ the total magnetic helicity is small, while
$|H_m/\tilde{H}_m|\sim 1$ is an indication that there is a
significant magnetic helicity. In Fig.~\ref{fig:NormHmag} (top panel) 
we show the evolution of the normalized helicity for the different
initial magnetic field strengths. Note that in all cases $|H_m/\tilde{H}_m|$ grows 
in time, and around 60 ms it has reached a substantial fraction of 1.
This fraction is higher for weaker fields, where the
evolution is smoother, \ie $|H_m/\tilde{H}_m| \sim 0.6$ for $B_{1.5}$,
while $|H_m/\tilde{H}_m| \sim 0.3$ for $B_{6.5}$. This behaviour
supports the idea that configurations with a significant amount of 
magnetic helicity are more stable, as already suggested in the literature
[see \citet{BraitSpruit06} and references therein]. 
Such indication is compatible with the observed instability of the initial 
purely poloidal field and the production of a toroidal component, which 
could be interpreted as an attempt of the system to produce magnetic helicity.

We have seen that the system can achieve a significant amount of 
magnetic helicity, but in the meanwhile most of its magnetic energy is lost. 
In order to weigh the absolute variation of magnetic helicity, we can 
compare $H_m$ with the initial value of $\tilde{H}_m$ (bottom panel 
of Fig.~\ref{fig:NormHmag}).
The ratio in this case is much smaller, $\lesssim 1-3\%$, indicating that the 
magnetic helicity produced would represent a small amount for the initial star, 
and becomes significant only because the system has lost most of its magnetic 
energy.  

On the basis of these results and bearing in mind the limits of our approach, 
we conjecture that magnetic helicity could play 
an important role as a stabilizing element for a fluid magnetized star. 
If this is the case, we can give a natural 
interpretation of the evolution of our system. (i) A purely poloidal magnetic field 
is unstable and a significant modification of its topology is necessary for 
stabilization. (ii) While the system is evolving the rate of change of magnetic 
helicity is small compared to the one of magnetic energy, thus the system has 
to lose most of its magnetic energy in order to significantly alter its topology 
and reach a more stable state.

As a complement to the information obtained from our nonlinear
relativistic calculations, we should remark that
the possibility of a stable equilibrium has been questioned in the
case of a barotropic fluid star \citep{Reisenegger2009,Lander2012}. 
A barotropic equation of state does not take into account the effect of
stable stratification associated to composition gradients, which could
play an important role in determining the stability of magnetic
equilibria in relativistic NSs~\citep{Reisenegger1992,
  Reisenegger2009}. Stable stratification, indeed, offers an
additional degree of freedom which favours the hydrostatic balance of
fluid and magnetic forces. A strong indication in support of
stable equilibria in stratified magnetized stars has been obtained for
main-sequence stars, where stratification is provided by entropy
gradients \citep{BraithNord2006}, while there is no such evidence for 
barotropic NSs.

\begin{figure*}
  \begin{center}
     \vskip 0.3cm
      \includegraphics[angle=0,width=5.8cm]{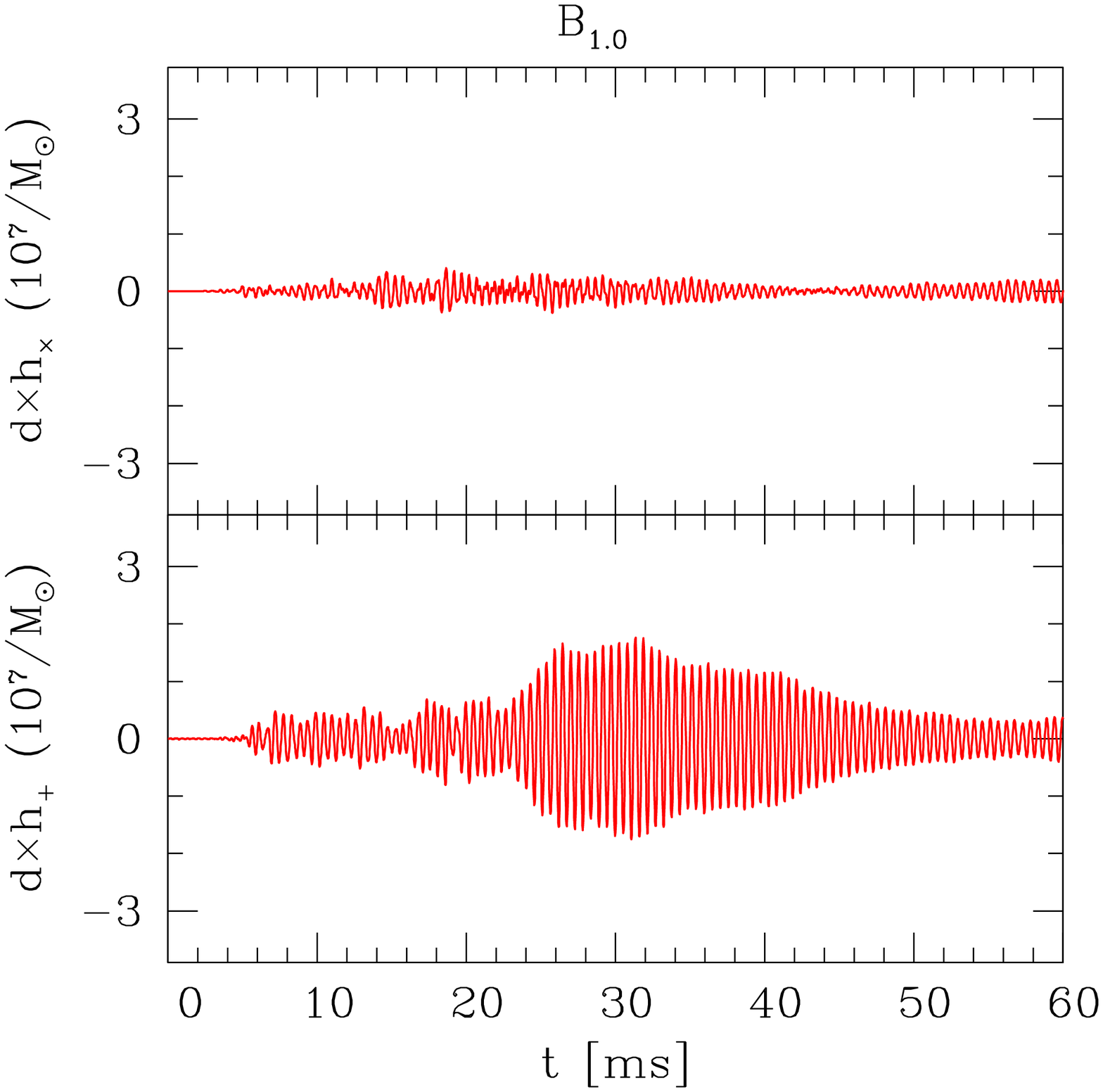}
     \hskip 0.1cm
     \includegraphics[angle=0,width=5.8cm]{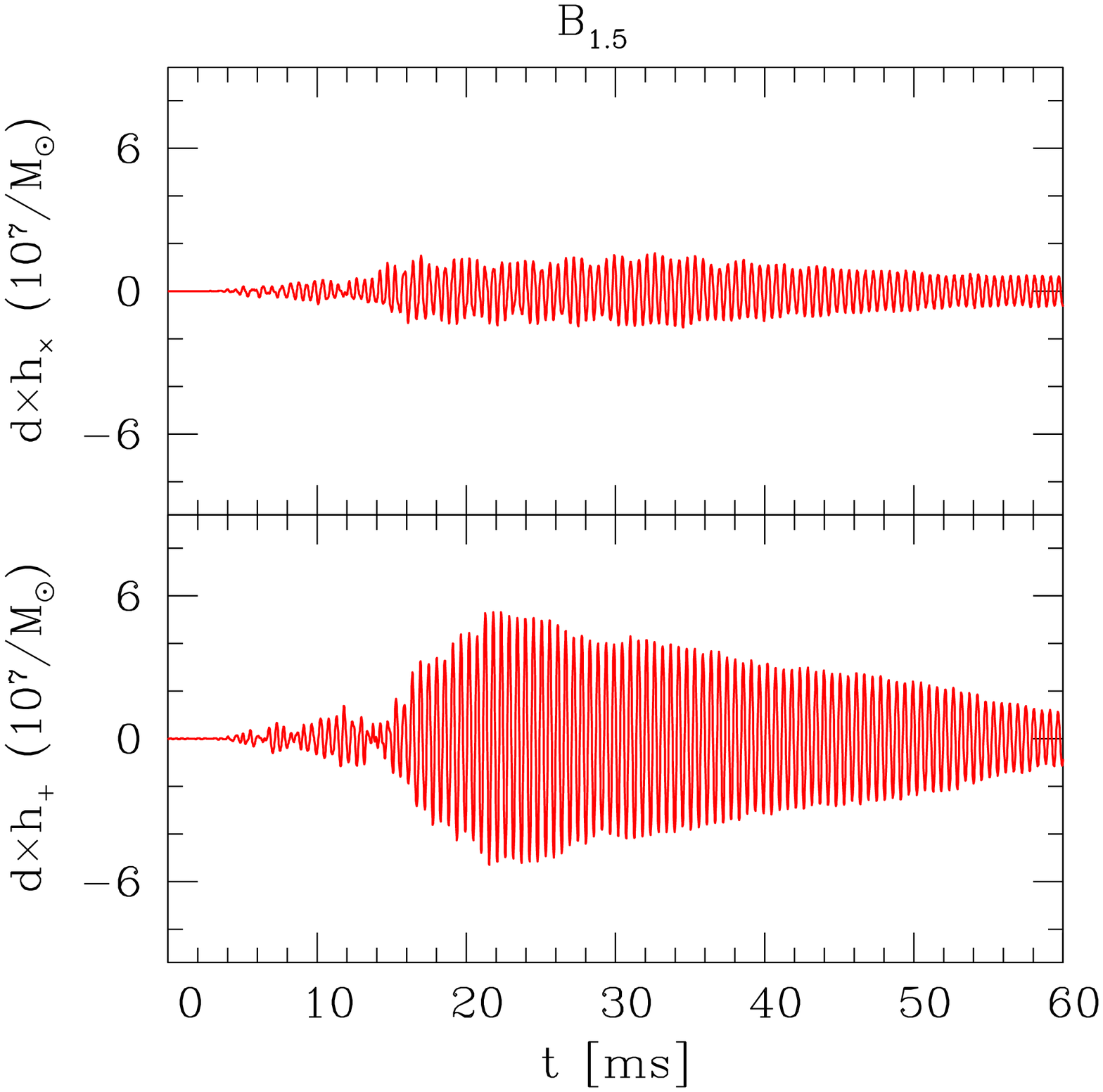}
     \hskip 0.1cm
     \includegraphics[angle=0,width=5.8cm]{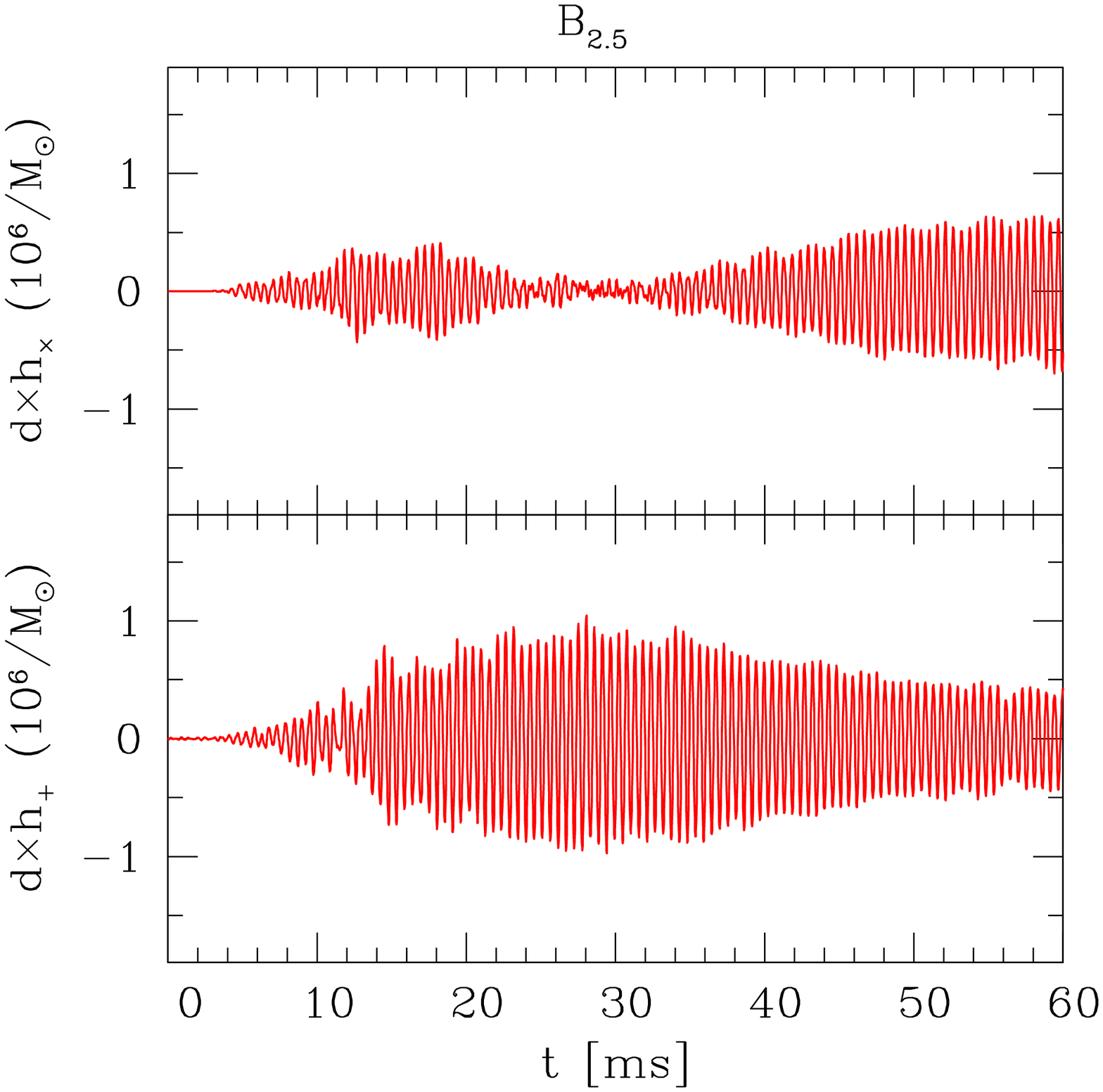}
     \vskip 0.3cm
     \includegraphics[angle=0,width=5.8cm]{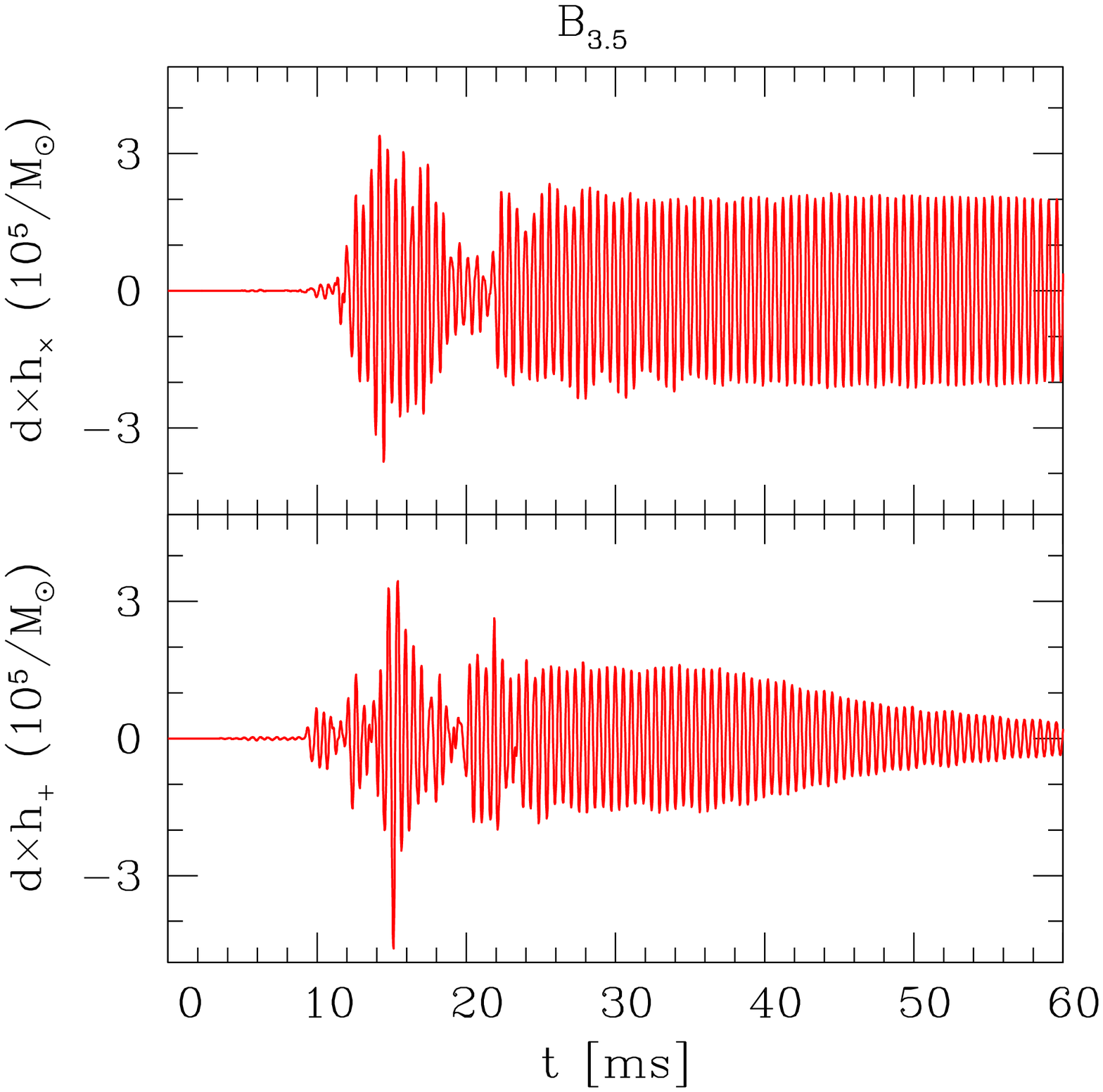}
     \hskip 0.1cm
     \includegraphics[angle=0,width=5.8cm]{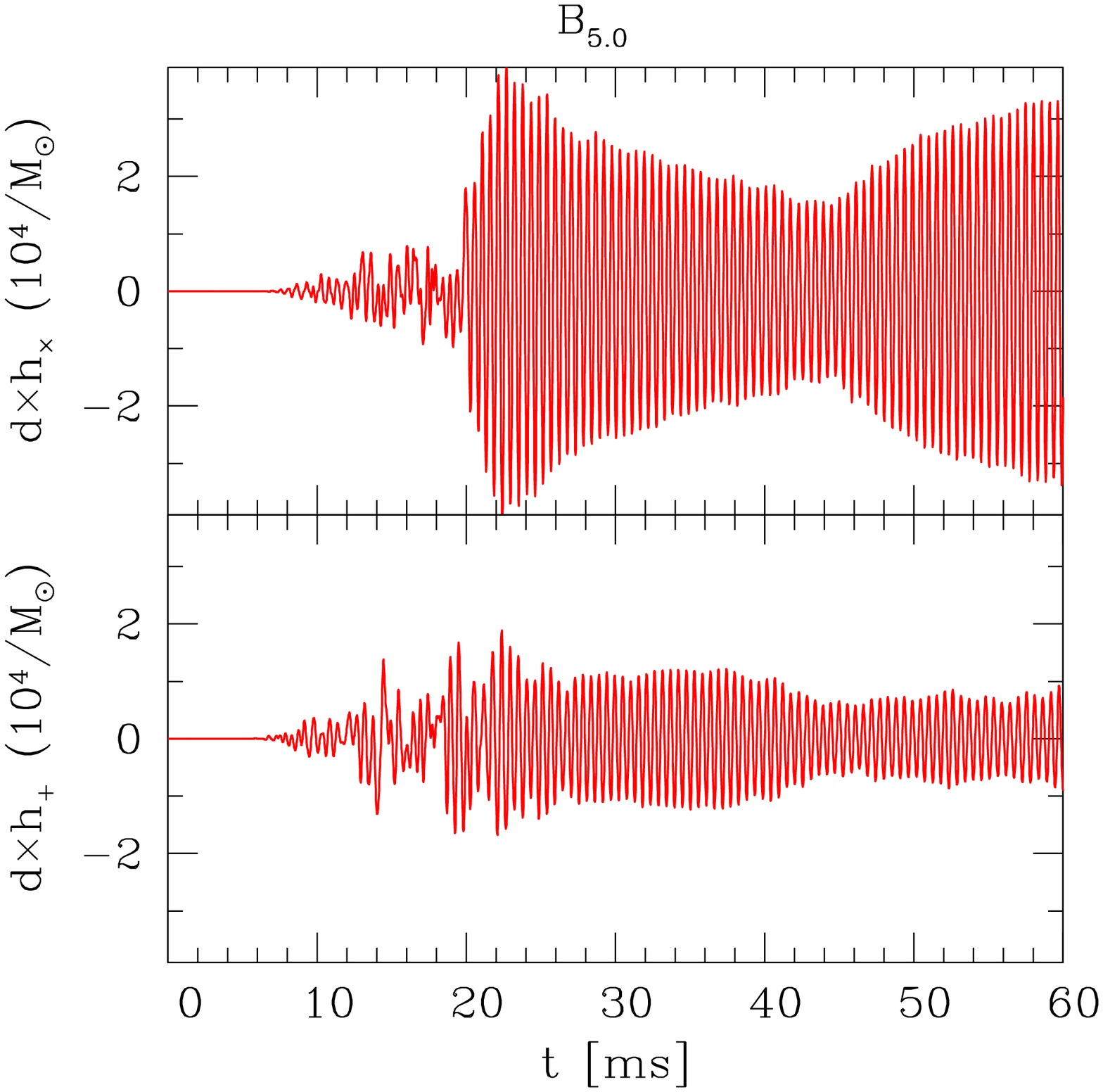}
     \hskip 0.1cm
     \includegraphics[angle=0,width=5.8cm]{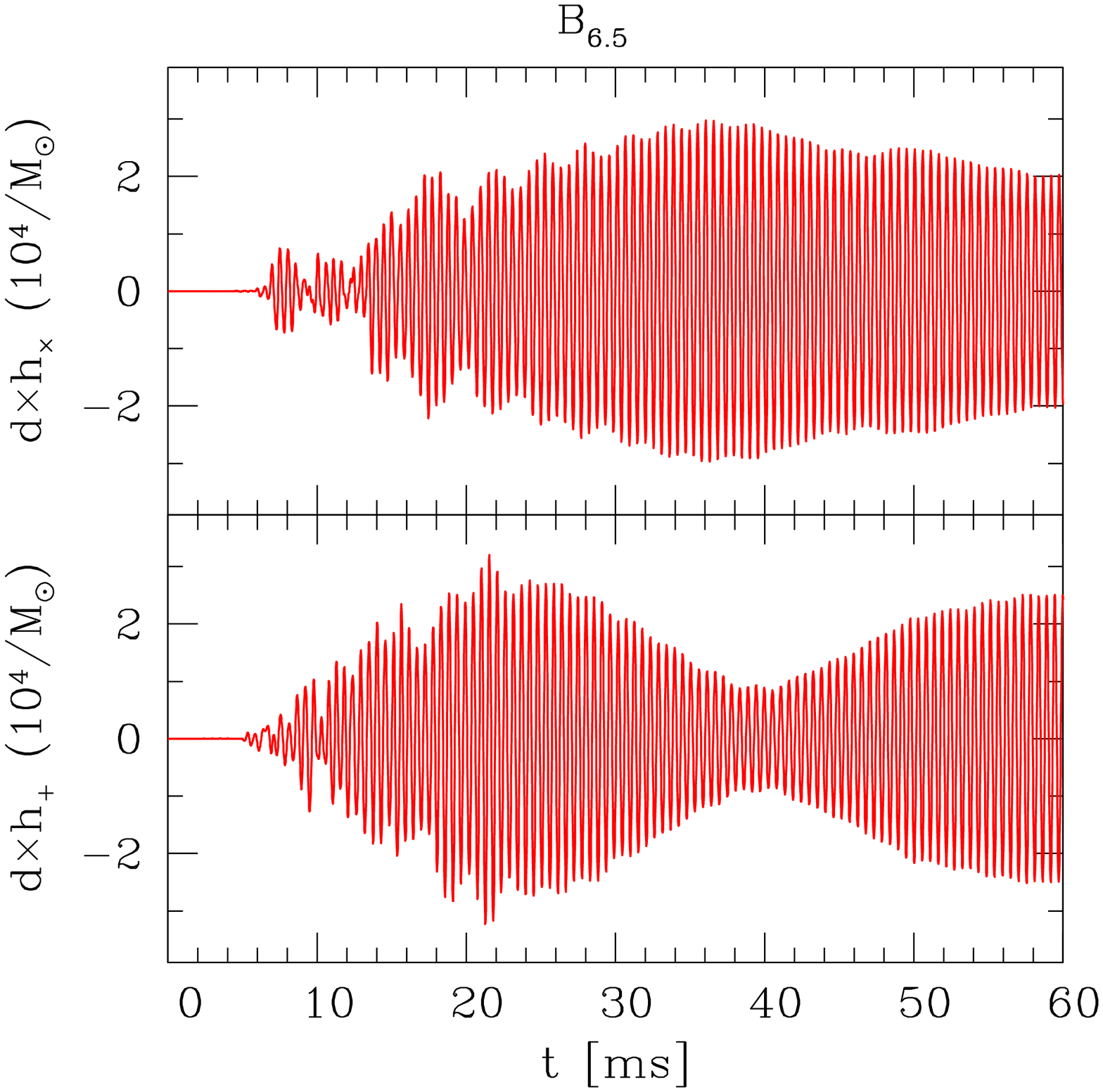}
  \end{center} 
  \caption{Gravitational wave signal for different initial magnetic
    field strength $B_{\rm p}$, in the range $B_{\rm p} =
    B_{1.0}-B_{6.5}$. Top (bottom) panels refer to the amplitude in
    $\times$ ($+$) polarization; $d$ is the source distance, which
    needs to be specified to obtain the amplitudes $h_\times$,
    $h_+$. Note that the $y$-scale changes with $B_{\rm p}$.}
\label{fig:GWsignal}
\end{figure*}


\subsection{Electromagnetic emission}
\label{ememission}

As discussed in the Introduction, one of the leading explanations of
magnetar giant flares assumes that the event is due to a sudden, global
rearrangement of the internal magnetic field, \ie the same kind of
process that we are studying. A giant flare starts with an initial
burst of $0.1-0.5\,{\rm s}$ duration, followed by a long pulsating
tail lasting hundreds of seconds. In such scenario, the energy emitted
in the initial burst is a significant fraction of the total magnetic
energy of the star, while the duration of the burst is dictated by the
timescale of the internal field rearrangement. We have also remarked
repeatedly that in all cases considered, our magnetized stars lose
most of the initial magnetic energy within few tens of ms. This energy
is converted in very small part into fluid motions, \eg $f$-mode oscillations of
the star with consequent emission of gravitational waves (see
Sect.~\ref{gravitationalwaves}), while most of it is dissipated in the
resistive layer inside the star and then in the atmosphere, thus
mimicking the radiative losses that would be measured if the stars
were actually in vacuum. We can therefore use the information about
the dissipated magnetic energy to deduce, within the approximation of
our approach, an order-of-magnitude estimate of the electromagnetic
luminosity $L_{em}$ associated with the hydromagnetic instability. In
addition, we can compare the luminosity and duration of the emission
computed in this way with the observations of giant flares.

In practice, most of the emission comes from the initial fast drop of
magnetic energy by $\sim 90\%$, corresponding to an initial spike in
$L_{em}$. Taking that as a reference, we measure the duration of the
spike and compute its peak and (time) average luminosities for the
different initial magnetic field strengths considered. Note that the
expected scaling for the luminosity is $L_{em}\propto B_{\rm
  p}^{\,3}$, since the timescale of the emission goes as $1/B$, while
the magnetic energy scales as $E_m\propto B^2$. Our estimate for the
average luminosity, obtained by fitting for the different magnetic
field strengths, gives
\begin{equation}
\label{emlum}
\langle L_{em} \rangle= 1.9
\times\left(
\frac{B_{\rm p}}{10^{15}~{\text G}}\right)^3\times 10^{48}\,{\rm erg/s} \,.
\end{equation}
Similarly, for the reference value of a magnetar-type magnetic field of
$B_{\rm p} = 10^{15}\,{\rm G}$, we obtain a duration of the initial spike 
of  $\sim 0.7 \,{\rm s}$ and a peak luminosity of 
$L_{em,peak} \sim 5.3 \times 10^{48}\,{\rm erg/s}$.

We can now compare with the observations. In the brightest giant flare
detected, the one in 2004 from SGR 1806-20, the initial spike lasted
$\sim 0.5\,{\rm s}$, had a peak luminosity of $\sim (2-5)\times10^{47}
\times (d/15\,{\rm kpc})^2\,{\rm erg/s}$, where $d$ is the distance of
the source, and accounting for the total isotropic energy radiated
(which was $\sim 99$\% of the total energy emitted in the whole giant
flare event, including the $\sim 400$ s tail), the average luminosity
was $\sim (0.3-1) \times 10^{47} \times (d/15\,{\rm kpc})^2\,{\rm
  erg/s}$ [see \citet{Mereghetti2008} and references therein].  The
duration of the initial burst from SGR 1806-20 is clearly comparable
with our estimate. The difference in average luminosity (our estimate
is $\sim 20-60$ larger), then, essentially reflects a difference in
the electromagnetic energy released, which does not constitute a
significant limitation for the scenario: while in our system the
magnetic energy is almost completely lost, it is perfectly reasonable
that in a more realistic situation the magnetic field would migrate
from one configuration to another with a smaller jump in magnetic
energy.  Note that the timescale of the process, on the other hand, is
essentially controlled by the initial internal magnetic field strength
and poorly depends on the overall jump in magnetic energy.
Interestingly, also the ratio between the estimated peak and average
luminosities is in good agreement with the observations.

If the comparison with the phenomenology of SGR 1806-20 leads to
a reasonably good match in the duration and energetics, thus providing
a substantial support to the internal field rearrangement scenario,
the comparison with other observations is not as striking. In
particular, the average luminosity of the initial spike in the case of the other two
giant flares observed (1979 from SGR 0526-66 and 1998 SGR 1900+14) was
$\lesssim10^{45}\,{\rm erg/s}$, with a comparable duration of
$\sim 0.25,0.35\,{\rm s}$. In these cases, therefore, the mismatch in
the energy losses is far larger, although still acceptable if the
field rearrangement is to a new configuration with comparable magnetic
field strengths.

The luminosity rise time at the beginning of the initial spike constitutes
an additional relevant timescale in a giant flare. Being of the order of 
$\sim 1$~ms \citep{Palmer2005}, it can be easily associated with the 
Alfv\'en propagation time in the magnetosphere, while it can be hardly 
explained by the internal magnetic field rearrangement, which acts on 
longer timescales. Therefore, in order to have the internal scenario 
fully compatible with the observations one probably needs to assume
that the instability also triggers an initial, less energetic but sudden 
reorganization of the magnetospheric field, with a mechanism similar 
to the one proposed in the alternative external rearrangement scenario 
\citep{Lyutikov2003,Lyutikov2006,Gill2010}. 
However, this feature can not be  captured by our present modelling 
of the NS exterior. 

In summary, although the dynamics of our system is oversimplified when
compared to the complex phenomenology shown by a realistic magnetar
giant flare, the overall agreement in the duration of the burst and in
its energetics when compared with the observations of the giant flare
from SGR 1806-20, lends support to the suggestion that the basic
phenomenology is that of an internal-field readjustment. To the best
of our knowledge, this is the first time that a comparison between 
general-relativistic MHD simulations and the magnetar phenomenology
has been attempted. Clearly, a lot more work needs to be done to
improve our self-consistent but crude modelling.


\subsection{Gravitational wave emission}\label{gravitationalwaves}

In the Introduction we have pointed out the potential relevance of our
study for GW observations in connection to magnetar giant flares. 
The basic idea, already presented in \citet{Ciolfi2011}, is that the 
instability triggers stellar oscillations at the $f$-mode 
frequency, with the consequent emission of GWs, similarly to what 
should happen in association with a magnetar giant flare.
In what follows we provide a systematic assessment of
this emission for the different initial magnetic field strengths
considered and an estimate of the detectability of the GW signal for
the planned advanced GW detectors, \ie advanced LIGO and advanced
Virgo, and the new generation ones, \eg Einstein
Telescope~\citep{Punturo2010}.

The GW signals produced in our simulations are shown in the different
panels of Fig.~\ref{fig:GWsignal}, where we report the amplitudes in
the $\times$ and $+$ polarizations for the various initial
magnetic field strengths $B_{\rm p}$, ranging from $B_{1.0}$ to $B_{6.5}$. Besides the
differences in the overall amplitude, that naturally grows with $B_{\rm p}$
(see also discussion below), the waveforms show also other differences
with the magnetic field strength. These include variations in the
transient stage of the instability and low-frequency modulations. For
example, for $B_{6.5}$ we can observe a significant modulation at
$\sim 25\,{\rm Hz}$, which could be associated with the jump in
central density (\cf Fig.~\ref{fig:rhomax}), whose timescale is
compatible with the period of the modulation. A similar modulation
appears also for $B_{2.5}$. Rather irregular transients characterize
instead the initial amplitudes for $B_{3.5}$ and $B_{5.0}$. In both
cases, in fact, we can notice sudden jumps in the signal, which
correspond to the highly dynamical phases in the evolution of the
system. This is indeed confirmed by the evolution of the total magnetic
energy in Fig.~\ref{fig:energies}, where bumps appear around
$\sim 12\,{\rm ms}$ and $\sim 22\,{\rm ms}$ for $B_{3.5}$ and $B_{5.0}$,
respectively, while they are not present in the other cases.

\begin{figure}
  \begin{center}
     \includegraphics[angle=0,width=8.0cm]{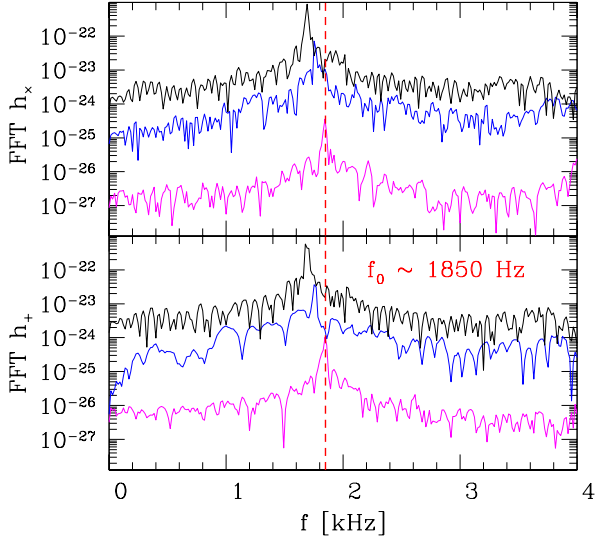}
  \end{center} 
  \caption{Fourier transform of $h_{\times}$ and $h_{+}$ for
    $B_{\rm p}=B_{1.5}$ (lower magenta line), $B_{3.5}$ 
    (intermediate blue line), and $B_{6.5}$ (upper black line), 
    assuming a source distance of 10 kpc. The vertical line marks a frequency of 
    $1850\,{\rm Hz}$.}
  \label{fig:fft}
\end{figure}

Despite these differences, all signals are essentially dominated by
oscillations at the $f$-mode frequency, as it is clear from
Fig.~\ref{fig:fft}, where we report the Fourier transform of the
GW amplitude in both the $\times$ and $+$ polarization for $B_{1.5},
B_{3.5}$ and $B_{6.5}$. The $f$-mode we can read off the plot is about
$1.85\,{\rm kHz}$ for the lowest field strength, with a shift to lower
frequencies for higher magnetic fields (this shift is of $\sim 8\%$
for $B_{6.5}$). We recall that since our evolutions are in the Cowling
approximation, the corresponding $f$-mode frequencies are notoriously
larger of $\sim 15\%$ than the correct ones computed in full general
relativity~\citep{Dimmelmeier2006,Takami2011}. This property will be
taken into account when considering the detectability of the signal.

Given the complexity of the signal, with large rapid and secular
variations, the determination of the signal-to-noise ratio (SNR)
cannot be done by simply looking at the maximum-minimum amplitudes,
but rather by computing a strain which represents a suitable time
average. This is indeed what can be done through the root-sum-square
amplitude $h_{\rm rss}$, defined as
\begin{equation}
h_{\rm rss} \equiv 
\left[ \int_{-\infty}^{+\infty}  \, h(t)^2\, dt \right]^{1/2} \,,
\label{eqhrss}
\end{equation}
where $h^2(t)=h_{\times}^2(t)+h_{+}^2(t)$. The amplitudes in the
two polarizations are computed considering a source within the Galaxy,
\ie at a fiducial distance of $10\,{\rm kpc}$. If, as in our case, the
signal is dominated by a single frequency $f_0$, then the SNR can be
estimated using the simple expression, $S/N=h_{\rm
  rss}/\sqrt{S_h(f_0)}$, where $\sqrt{S_h(f_0)}$ is the strain-noise
amplitude of the detector at the frequency $f_0$. When computing
$h_{\rm rss}$ we also need to specify the duration of the signal (in 
addition to the distance of the source). 
Typical estimates of the $f$-mode damping time
$\tau_{_{GW}}$ range between $100\,{\rm ms}$ and $1\,{\rm
  s}$~\citep{Andersson1998,Benhar2004}, and here we assume
$\tau_{_{GW}}=100\,{\rm ms}$.

\begin{figure}
  \begin{center}
  \includegraphics[angle=0,width=8.0cm]{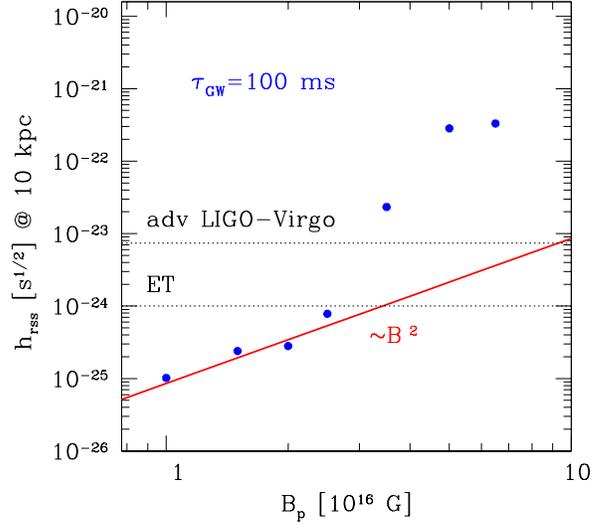}
  \end{center} 
  \caption{$h_{\rm rss}$ $[\sqrt{s}]$ versus $B_{\rm p}$ in Gauss assuming a
    distance of $10\,{\rm kpc}$ and a damping time of
    $\tau_{_{GW}}=100$~ms. Horizontal lines
    mark the strain-noise amplitude of the considered GW detectors at
    the $f$-mode frequency $f_0$. Finally, the red line is obtained by
    imposing a quadratic scaling ($\propto B^2$) and fitting the
    results for the lowest magnetic fields (first four points).}
  \label{fig:gw_scaling}
\end{figure}

The resulting $h_{\rm rss}$ amplitudes are reported in
Fig.~\ref{fig:gw_scaling}, where we also show with dotted lines the
strain-noise amplitude of advanced LIGO and Virgo and of the Einstein
Telescope, and where we have assumed $f_0\sim 1550\,{\rm Hz}$ to
correct for the Cowling approximation. Furthermore, the ``noise'' is
estimated to be $\sqrt{S_h(f_0)}\sim 7.4 \times 10^{-24}\,{\rm
  Hz}^{-1/2}$ for advanced LIGO and and advanced Virgo, and
$\sqrt{S_h(f_0)}\sim 10^{-24}\,{\rm Hz}^{-1/2}$ for the Einstein
Telescope [see \citet{andersson2011}, where the same strain-noise
  amplitudes were considered].

Because in our simulations we have considered magnetic-field strengths
that are at least an order of magnitude larger than those typically
associated with magnetars, we need to rescale our results for $h_{\rm
  rss}$ to smaller field strengths in order to assess the GW signal
that could be expected if the instability is associated to a magnetar
giant flare. This is rather straightforward to do since the
expectation from perturbative analyses [see, \eg, \citet{Levin:2011}]
is that, independently on the mechanism considered, the amplitude of
the excited $f$-mode by the flare should scale as the magnetic energy,
\ie as $\propto B^2$. This prediction is valid as long as the magnetic
field strengths are sufficiently small so that the corresponding
magnetic energy can be considered as a small perturbation to the total
binding energy. The results shown in Fig.~\ref{fig:gw_scaling} confirm
that a quadratic scaling is a rather good approximation to the
computed data for $B_{\rm p} \lesssim B_{3.0}$. This is highlighted in
Fig~\ref{fig:gw_scaling} by the red solid line, which represents a
quadratic fit to the GW amplitude in the case of low-magnetic
fields. We find it very reassuring that our data does show the
expected scaling behaviour in the weak-field limit, which instead
appears to be absent in the scaling reported by~\cite{Zink:2011} and
subsequently by~\cite{Lasky2012} for rotating stars. Of course, this
scaling is then broken as the magnetic field is increased and a
steeper dependence is then expected and found for $B_{\rm p} \gtrsim
B_{3.5}$. The characteristic amplitude then appears to
saturate for $B_{\rm p} \gtrsim B_{5.0}$. We note that the
very rapid increase in $h_{\rm rss}$ for $B_{\rm p} \simeq
B_{3.5}$ is somewhat surprising but not completely unexpected. 
We recall, in fact, that the root-sum-square amplitude refers to a
secondary quantity which is calculated in a Newtonian
approximation. Because of the high time-derivatives of the fluid
variables that contribute to its measure, this quantity is very
sensitive to the dynamics of the system and hence to the large
velocities that develop for large magnetic fields. By using different
prescriptions of the resistivity we have verified that this rapid 
change is robust, although probably amplified by the breaking of 
the Newtonian approximation.

The importance of the quadratic fit is that it allows us to extrapolate 
back to even smaller values of the magnetic fields and obtain the 
following predictions for the SNRs for the different detectors:
\begin{align}
\left.\frac{S}{N}\right|_{\rm AdvLIGO-Virgo} & \simeq 1.2 \times 10^{-4}
\times\left(\frac{B_{\rm p}}{10^{15} \;\text{G}}\right)^2\,,\\
\left.\frac{S}{N}\right|_{\rm ET} & \simeq 0.9 \times 10^{-3}
\times\left(\frac{B_{\rm p}}{10^{15} \;\text{G}}\right)^2 \,.
\label{SNformula}
\end{align}
We conclude that the GW signal produced by the $f$-mode oscillations
triggered by the hydromagnetic instability in association with a magnetar 
giant flare will be undetectable for realistic values of the magnetic field, 
\ie $B_{\rm p} \lesssim 10^{15}\,{\rm G}$ and marginally detectable by 
third generation detectors for unrealistic magnetic fields, 
$B_{\rm p} \gtrsim 2\times 10^{16}\,{\rm G}$. Because a longer damping 
time, say of $\tau_{_{GW}}=1\,{\rm s}$, would yield a gain of only a factor
$\sqrt{10}$, we do not expect that the prospects of a detection would
improve if the $f$-mode oscillations would last a factor ten more in
time.

Understanding why the detectability of this GW signal is so hard in
practice can be made easier if we take into account how much energy is
actually lost to gravitational waves and compare it with the amount of
dissipated magnetic energy. More specifically, if $h_{\rm rss}$ is the
root-sum-square GW amplitude, the energy emitted in GWs can be
computed as [see, \eg \citet{Levin:2011}]
\begin{align}
E_{GW} & = \frac{2\pi^2 f_0^2 c^3}{G} d^2 h_{\rm rss}^2  \,, 
\nonumber \\
& \simeq 1.9 \times 10^{37} \;\text{erg} \times
\left(\frac{B_{\rm p}}{10^{15} \;\text{G}}\right)^4
\times\left(\frac{\tau_{_{GW}}}{100\text{~ms}}\right) \,,
\label{GWfraction}
\end{align}
where we retained factors of $c$ and $G$, and $d$ is the source
distance\footnote{Note that $d^2 h_{\rm rss}^2$ does not depend 
on the source distance, while $h_{\rm rss}$ alone does.}. 
Comparing expression~\eqref{GWfraction} with the
corresponding estimates of the dissipated magnetic
energy~\eqref{emlum}, it is easy to realize that the two energies
differ by more than ten orders of magnitude. Hence, assuming
$E_{GW}\sim E_m$ as done recently by \citet{Corsi2011} would indeed
lead to a $S/N \sim 30$ for advanced LIGO, but it would also be overly
optimistic and rather unjustified on the basis of our calculations, which 
instead confirm the expectation of \citet{Levin:2011} that the energy 
actually converted into GWs is only a small fraction of the total energy 
available.
As a result of this inefficient conversion, the GW signal is too weak to be 
detected with the near future detectors. A similar conclusion was also reached by
\citet{Levin:2011,Zink:2011}, and more recently by \citet{Lasky2012} for 
rotating stars.


\section{Summary and conclusion}
\label{sec:conclusions}

We have performed 3D general-relativistic MHD simulations of
nonrotating magnetized NSs endowed with a purely poloidal magnetic
field and studied the development of the hydromagnetic instability
that develops dynamically. This work represents an extension of our
previous study on the subject \citep{Ciolfi2011}, where we have
improved our treatment of the atmosphere, drastically reducing the
undesired energy losses due to the transition between the ideally
conducting stellar interior and the resistive exterior, and we have
performed simulations on much longer timescales, which have
allowed us to gain essential information about the final configuration
reached by the system. The resulting overall picture is much clearer
and conclusive.

As expected from perturbation analyses, the instability is first
triggered in the region of close field lines and is accompanied by the
production of a toroidal magnetic field. When the growth of the latter
saturates in about one Alfv\'en timescale, the toroidal field has
reached a local strength which is comparable to the poloidal one and
the axisymmetry of the initial configuration is lost. At this point,
the most dynamical phase of the nonlinear evolution takes place, with
major modifications of the magnetic field leading to a strong
electromagnetic emission carrying away $\sim 90\%$ of the magnetic
energy in few Alfv\'en timescales. At the same time, a small fraction
of magnetic energy is converted into stellar oscillations, mostly at
the $f$-mode frequency, which cause the emission of gravitational
waves. The subsequent evolution proceeds on longer timescales, with
further loss of magnetic energy.

Since in the post-instability phase the magnetic field is continuously
changing, losing strength because of resistive dissipation, it is
difficult to determine whether the corresponding configuration is a
stationary one or not. The only robust evidence is that the variations in the
hydrodynamical and electromagnetic quantities are much smaller than
those during the instability, taking place on larger timescales. We
therefore interpret this behaviour as evidence that the new
magnetic-field configuration has reached a quasi-stationary state or
that, if still intrinsically unstable, the growth time of the
instability is much larger than the one that can be possibly
investigated numerically.

Because in this quasi-stationary configuration the ratio of the
toroidal and poloidal magnetic energies tends to unity, and because
the growth of toroidal magnetic field at the expense of the poloidal
one is also accompanied by an increase in the magnetic helicity, we
are led to conclude that an equilibrium magnetic field configuration
with a significant amount of magnetic helicity and comparable poloidal
and toroidal magnetic field energies could be a preferred one for
stability. Bearing this in mind, we remark that it is still unclear if
stable equilibria exist for a simple barotropic fluid star, and that
other stabilizing contributions, such as a stable stratification, may
be necessary to obtain long-lived magnetic field configurations.

The violent reorganization of magnetic fields induced by the
development of a hydromagnetic instability has been proposed as a
possible mechanism to explain giant flares in magnetars. Using our
simulations and in particular the information about the dissipated
magnetic energy, we have deduced, within the approximation of our
resistive approach, an estimate of the electromagnetic luminosity
$L_{em}$ associated with the hydromagnetic instability. More
specifically, we found that the average luminosity scales as $\propto
B^3$ with the magnetic-field strength, in good agreement with the
expectation that the radiated energy should scale as $\propto B^2$,
while the duration of the emission should scale as $\propto B^{-1}$.
In this way we were able to perform a direct comparison with the
observations of SGR 1806-20, finding a very good agreement with the
duration of the burst and an emitted luminosity which is about an
order of magnitude larger than the measured one. As a result, although
our modelling is oversimplified, the overall agreement with the
observations from SGR 1806-20, lends support to the suggestion that
the basic phenomenology is that of an internal-field readjustment.

Finally, we have discussed the gravitational-wave emission which
should be expected as a result of the $f$-mode oscillations triggered by
the instability. Also in this case, our calculations reproduce the
expected scaling behaviour of the root-mean-square gravitational-wave 
amplitude in the limit of weak magnetic fields, \ie $h_{\rm rss} \sim
B^2$. This important result allows us to extrapolate with confidence
our estimates to smaller magnetic fields than those covered by our
simulations and conclude that the gravitational-wave signal from
$f$-mode oscillations will be undetectable for realistic values of the
magnetic field, \ie $B_{\rm p} \lesssim 10^{15}\,{\rm G}$ and marginally 
detectable by third generation detectors for unrealistically large magnetic
fields, \ie $B_{\rm p} \gtrsim 2\times 10^{16}\,{\rm G}$. 

As a self-consistent solution to this problem in fully resistive
regime is close to be within reach~\citep{Kiki2012}, we plan to extend
the investigation reported above and thus remove many of the
approximations that our present analysis had to sustain. It will then
be possible to set even more precise connections between the complex
phenomenology observed in magnetar flares and the violent dynamics of
hydromagnetic instabilities in relativistic stars.

\bigskip
We are grateful to D. Radice, A. Harte, S.~K. Lander, S. Mereghetti, 
W. Kastaun, B. Giacomazzo, E. Bentivegna, G.~M. Manca, R. De Pietri 
and S. Bernuzzi for useful discussions. R. Ciolfi is supported by the
Humboldt Foundation. Support comes also from ``CompStar'', a Research
Networking Programme of the European Science Foundation and by the DFG
grant SFB/Transregio~7. The calculations have been performed on the
supercomputing clusters at the AEI.




\begin{figure}[h]
  \begin{center}
     \includegraphics[angle=0,width=8.0cm]{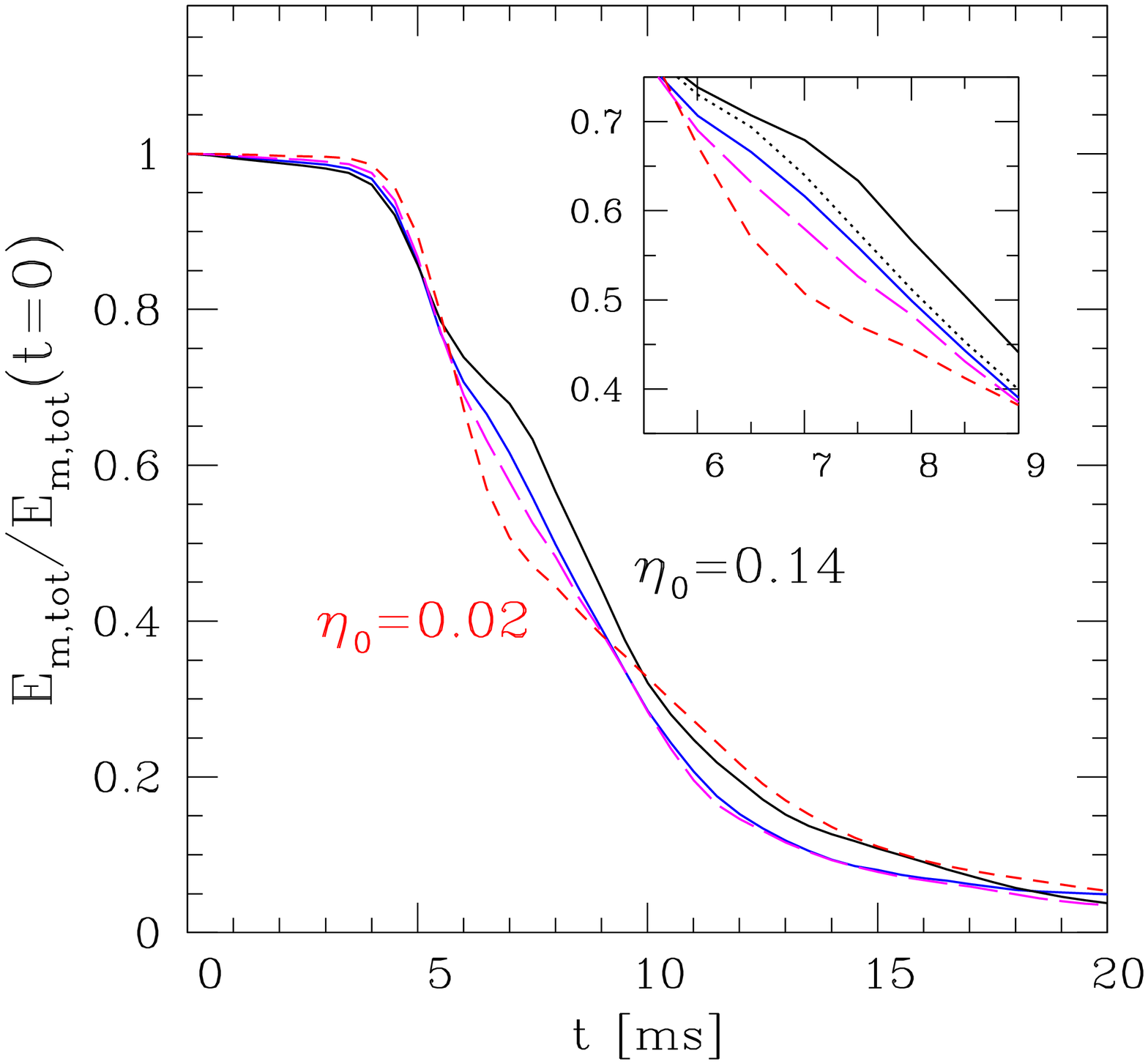}
     \hskip 0.4cm
     \includegraphics[angle=0,width=8.0cm]{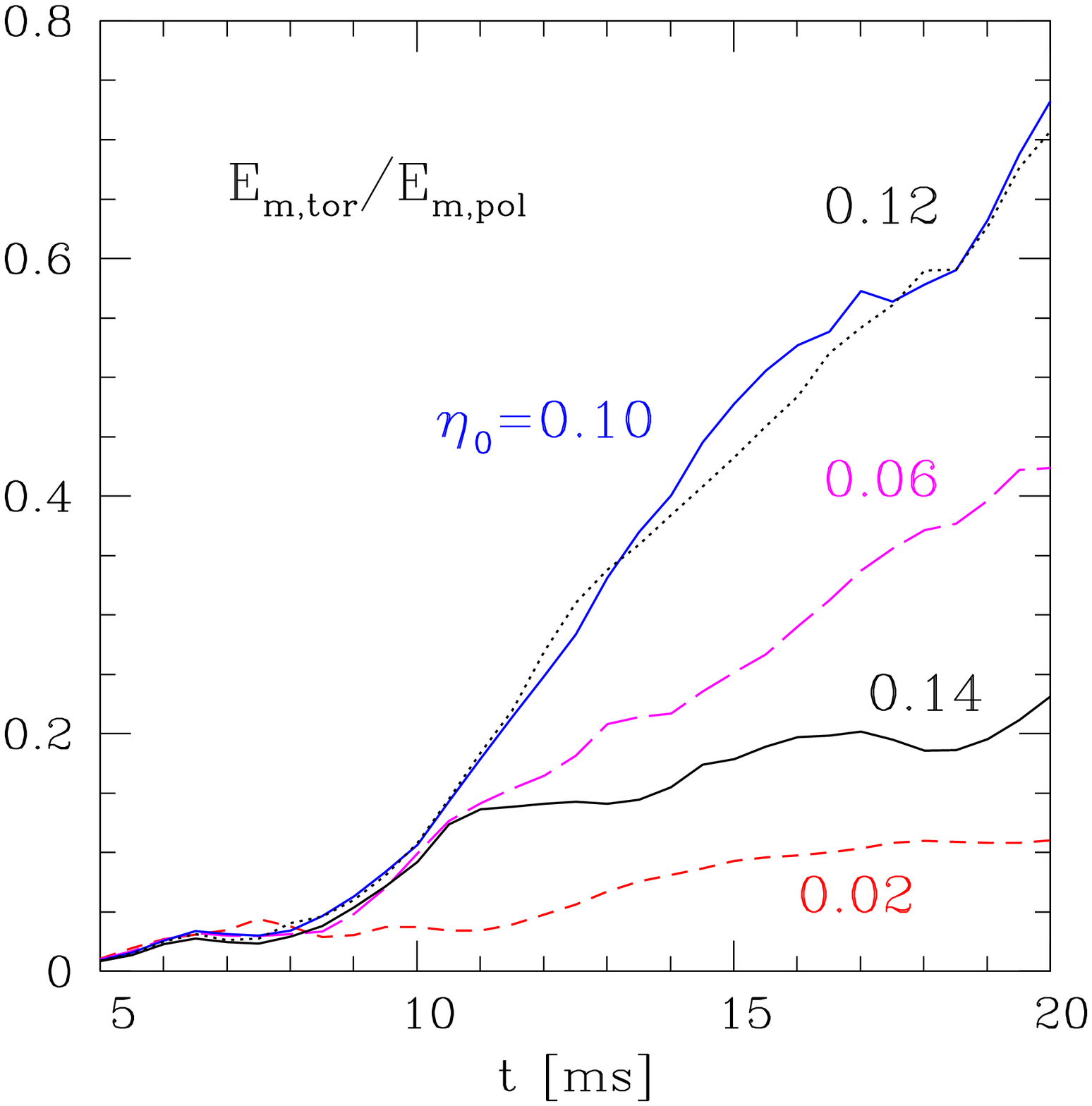}
  \end{center} 
  \caption{ {\it Left panel}. Total magnetic energy (normalized to the
    initial value) versus time for simulations with $B_{\rm p}=B_{6.5}$ 
    and different values of resistivity: $\eta_0/M_\odot=0.02$
    (dashed red), $0.06$ (long-dashed purple), $0.10$ (blue), $0.14$
    (black). In the insert we have also $\eta_0/M_\odot=0.12$ (dotted
    black).  {\it Right panel}. Ratio of toroidal and poloidal
    magnetic energies versus time for the same collection of
    simulations.  }
\label{fig:etadep}
\end{figure}

\appendix
\section{The role of resistivity in the atmosphere}
\label{etadepend}

In this Appendix we discuss how the results of our simulations depend
on the atmospheric value of resistivity $\eta_0$ and how a suitable
choice for $\eta_0$ can considerably limit the influence of this free
parameter. We recall that if the value of $\eta_0$ is too small, we
expect that the exterior field is not evolving rapidly enough relative
to the interior dynamics, thus accumulating magnetic field distortions
in the external layer of the star. As $\eta_0$ is increased, the
timescale for the exterior field evolution decreases until it becomes
comparable or shorter than the timescale of the internal evolution. At
that point the overall dynamics is less sensitive to a further
increase of $\eta_0$. Finally, if we further increase $\eta_0$, it
eventually will become too large and the effects of resistivity in the
outer layer of the star will become dominant, significantly
influencing also the interior dynamics.

In Fig.~\ref{fig:etadep} we show the evolution of the total magnetic
energy and of the ratio of toroidal and poloidal energies, obtained
with different values of $\eta_0$ ranging from $0.02$ to
$0.14\,M_\odot$, with an initial magnetic field strength $B_{\rm
  p}=B_{6.5}$.  We first focus on the total magnetic energy (left
panel of Fig.~\ref{fig:etadep}) and note that in the range $0.02$ to
$0.10\,M_\odot$, the differences among different evolutions become
smaller and smaller as the value of $\eta_0$ is
increased. Furthremore, while a value of $\eta_0 = 0.12\,M_\odot$
still gives comparable results, an additional increase leads again to
significant differences. As a result, for these magnetic field
strengths, $\eta_0 \simeq 0.06-0.12\,M_\odot$ represents a reasonable
value for the resistivity.

Let us now consider the toroidal-to-poloidal energy ratio (right panel
of Fig.~\ref{fig:etadep}), which is particularly sensitive to the
value of resistivity. When considering $\eta_0$ ranging from
$0.02\,M_\odot$ to $0.10\,M_\odot$, we have a significant increase of
the toroidal field production. Around $0.10\,M_\odot$, however, the
ratio becomes almost independent of $\eta_0$, with no significant
differences for $\eta_0$ between $0.10$ and $0.12\,M_\odot$. However,
if we further increase $\eta_0$ above $0.12\,M_\odot$, we again find a
significant dependence, this time with a decrease in the production of
toroidal-field.

In conclusion, for $B_{\rm p}=B_{6.5}$ a resistivity value of
$\eta_0=0.10\,M_\odot$ yields optimal results and the scaling of this
value for different magnetic field strengths can be done simply as to
$\eta_0/M_\odot=0.10 \times (B_{\rm p}/B_{6.5}$). In the
limit of low magnetic field, a change in $B_{\rm p}$ accompanied by
this rescaling of $\eta_0$ would give exactly the same dynamics, only
rescaled in time.


\end{document}